\documentclass[twocolumn,showpacs,preprintnumbers,amsmath,amssymb]{revtex4}

\usepackage{graphicx}
\usepackage{dcolumn}
\usepackage{bm}

\def\fun#1#2{\lower3.6pt\vbox{\baselineskip0pt\lineskip.9pt
\ialign{$\mathsurround=0pt#1\hfil##\hfil$\crcr#2\crcr\sim\crcr}}}
\def\lap{\mathrel{\mathpalette\fun <}}
\def\gap{\mathrel{\mathpalette\fun >}}

\def\mass{{\cal M}}
\def\msun{{\mass_\odot}}

\def\beq{\begin{equation}}
\def\eeq{\end{equation}}

\def\mh{M_{\bullet}}

\def\tc{t_\mathrm{coh}}
\def\lc{L_\mathrm{coh}}
\def\trr{t_\mathrm{RR}}
\def\tnr{t_\mathrm{NR}}
\def\tm{t_\mathrm{M}}
\def\rm{r_\mathrm{capt}}
\def\lm{L_\mathrm{m}}

\def\eccm{e_\mathrm{m}}

\begin{document}


\title{Stellar Dynamics of Extreme-Mass-Ratio Inspirals}
\author{David Merritt}
 \email{merritt@astro.rit.edu}
\affiliation{Department of Physics and Center for Computational Relativity and Gravitation, Rochester Institute of Technology, Rochester, NY 14623}
\author{Tal Alexander}
\email{tal.alexander@weizmann.ac.il}
\affiliation{Faculty of Physics,Weizmann Institute of Science, POB 26, Rehovot, Israel}
\author{Seppo Mikkola}
\email{mikkola@utu.fi}
\affiliation{Tuorla Observatory, University of Turku, V\"ais\"al\"antie 20, Piikki\"o, Finland}
\author{Clifford M. Will}
\email{cmw@wuphys.wustl.edu}
\affiliation{McDonnell Center for the Space Sciences, Department of Physics, Washington University, St. Louis, MO 63130}
\date{\today}

\begin{abstract}
Inspiral of compact stellar remnants into massive black holes
(MBHs) is accompanied by the emission of gravitational waves at
frequencies that are potentially detectable by space-based interferometers.
Event rates computed from statistical (Fokker-Planck, Monte-Carlo)
approaches span a wide range due to uncertaintities about the
rate coefficients.
Here we present results from direct integration of the 
post-Newtonian $N$-body equations of motion descrbing 
dense clusters of compact stars around Schwarzschild MBHs.
These simulations embody an essentially exact (at the post-Newtonian
level) treatment of 
the interplay between stellar dynamical relaxation, relativistic
precession, and gravitational-wave energy loss.
The rate of capture of stars by the MBH is found to be 
greatly reduced by relativistic precession, which 
limits the ability of torques from the stellar potential
to change orbital angular momenta.
Penetration of this ``Schwarzschild barrier'' does occasionally occur, 
resulting in capture of stars onto orbits that gradually 
inspiral due to gravitational wave emission;
we discuss two mechanisms for barrier penetration and find
evidence for both in the simulations.
We derive an approximate formula for the capture rate, which
predicts that captures would be strongly disfavored from
orbits with semi-major axes below a certain value; 
this prediction, as well as the predicted rate,
are verified in the $N$-body integrations.
We discuss the implications of our results for the detection of 
extreme-mass-ratio inspirals from galactic nuclei with a range of
physical properties.
\end{abstract}

\pacs{Valid PACS appear here}
\maketitle

\section{\label{sec:intro} Introduction}

Compact stellar remnants -- white dwarfs, neutron stars, and 
stellar-mass black holes (BHs) -- can be
captured by massive black holes (MBHs) at the centers of galaxies
like the Milky Way, without suffering tidal disruption in the process.
Such extreme-mass-ratio inspirals (EMRIs) are a potential
source of low-frequency gravitational waves for
space-based gravitational wave (GW) interferometers
\cite{Danzmann-98,Hughes-03,Barack-04,AmaroSeoane-07}.
Capture orbits for EMRIs can be very eccentric \footnote{Capture onto more
nearly circular orbits could occur if the compact object
was originally a member of a bound pair; see Miller et al.,
Astrophys. J. Letts., {\bf 631}, L117 (2005).},
displaying extreme versions of relativistic precession.
Typical EMRIs will have low, instantaneous GW amplitudes,
but signals can potentially be observed over $\gap 10^5$ cycles as the compact objects
gradually lose energy and spiral in, allowing the signal-to-noise
ratio to be built up over time using matched filtering or other
techniques \cite{Gair-04,Wen-05,Cornish-05}.
Detailed information about the structure of spacetime is
encoded in the GW signal, permitting strong-field
tests of theories of gravity \cite{Ryan-95,Berti-05,Drasco-06}.


Predictions of the EMRI event rate span a wide range,
from $\sim 10^{-9}$ yr$^{-1}$ to $\sim 10^{-6}$ yr$^{-1}$ 
per galaxy \cite{Hils-95,SR-97,Ivanov-02,HA-06a,HA-06b,Freitag-06}.
There are two basic sources of uncertainty.
Only stars originating from tightly bound orbits,
$a\lap 10$ mpc (milliparsecs), can complete their inspiral 
without being scattered prematurely into the
MBH or onto a wider orbit.
But the number and distribution of stars and stellar remnants
at these radii is essentially unconstrained, even in the Milky Way
\cite{Merritt-10},
and estimates of the event rate must therefore be based on 
extrapolation of the stellar distribution observed 
on much larger scales, or on theoretical models.
In addition, the collisional dynamics of relativistic star clusters 
around MBHs are poorly understood.

This paper addresses the second source of uncertainty.
We present results from long-term ($10^6-10^7$ yr ), 
direct $N$-body simulations of clusters of compact stars around a MBH.
Relativistic corrections to the equations of motion are included
up to 2.5 post-Newtonian order \cite[][hereafter Paper I]{MAMW-10}.
These new simulations permit an essentially exact treatment of the 
interplay between stellar relaxation and GW emission,
avoiding the approximations that must be made in statistical 
(Fokker-Planck, Monte-Carlo) treatments.

In such a statistical treatment of EMRI inspiral, Hopman \&
Alexander \cite{HA-06b} have shown that the dynamical evolution leading to
capture on an inspiral orbit is driven by ``resonant relaxation''
\cite{Rauch-96} due to the residual torques from the
stellar background. They argued that relativistic, in-plane (Schwarzschild)
precession plays a critical role in suppressing the stellar torques on
eccentric orbits, thereby allowing the stars to follow
quasi-periodic EMRI inspiral orbits, rather than be strongly torqued
 into direct plunge orbits; the latter would produce
non-periodic, broad-band gravitational wave events that would be difficult
to detect.

The simulations presented here reveal that the interplay between 
Newtonian torques and relativistic precession not only limits the 
effectiveness of stellar relaxation before it
can drive stars into plunge orbits, but in fact creates a dynamical
barrier, the ``Schwarzschild barrier,'' which repels stars back
to less eccentric orbits, thereby strongly mediating the EMRI
rate. We develop a Hamiltonian model for the behavior of orbits near
this barrier and use it to identify two modes
by which stars can cross the barrier and become EMRIs.
Evidence for both modes of barrier penetration are found in the
$N$-body simulations.

In Sec.~\ref{sec:models} we summarize the computational techniques and
the $N$-body initial conditions.
In subsequent sections we present results from integrations in
which the PN terms are absent, or included only at the 
2.5PN (GW emission) level (Sec.~\ref{sec:seronetwo}); 
and in which all PN terms
up to and including 2.5PN are included (Sec.~\ref{sec:serthree}).
In Sec.~\ref{sec:serthree} we also present an extended discussion
of orbital dynamics near the Schwarzschild barrier based on
a Hamiltonian formulation.
Sec.~\ref{sec:discuss} discusses the implications of our results
for the rate of EMRI production in real galaxies and
Sec.~\ref{sec:conclude} sums up.

We confine ourselves in this paper to non-rotating, i.e. Schwarzschild,
MBHs.
The consequences of spin will be discussed in a subsequent paper.

\section{\label{sec:models}Models and Methods}

The $N$-body systems considered here consist of a single massive
particle, representing a massive black hole (MBH), and 50
lower-mass particles representing stellar remnants
(referred to below, interchangeably, as either BHs or stars).
Each BH particle had a mass $5\times 10^{-5}$ that of the MBH
particle.
If the latter is assigned a mass of 
\beq
\mh = 1\times 10^6\msun,
\eeq
the BH particles have masses of $m=50\msun$.
This value is somewhat larger than the predicted masses of the
BHs that form in stellar collapse, i.e. $10-20\msun$ \cite{Woosley-02}.
The choice for $m/\mh$ was motivated by the need to integrate the
$N$-body systems for a time of order the two-body relaxation
time, which scales as $m^{-2}$ for a system of fixed $N$.
Alternatively, if $m$ is set to $10\msun$,
$\mh=2\times 10^5\msun$; however we note that the existence of MBHs
with $\mh\lap 10^6\msun$ is speculative.

Unless otherwise stated, we adopt $\mh=1.0\times 10^6\msun$
below when quoting $N$-body results in physical units.
In most cases, the dynamical theory used to interpret the $N$-body
results will allow the event rates derived here to be scaled approximately
to systems of different $m$ and $\mh$.

The initial orbital elements of the BH particles were 
selected randomly from the distribution
\beq\label{eq:Nofae}
N(a,e^2)da de^2 = N_0 da de^2
\eeq
with $a$ and $e$ the semi-major axis and eccentricity
of the Keplerian orbit about  the MBH.
Eq.~(\ref{eq:Nofae}) corresponds to an isotropic
(in velocity) distribution with configuration-space density
\beq\label{eq:nofr}
n(r) = n_0\left(\frac{r}{r_0}\right)^{-2}.
\eeq
This is roughly the expected radial dependence for a relaxed
population around a MBH \cite{BW-77,Freitag-06,AH-09}.
The initial distribution in semi-major axis was truncated above 
$a=a_2=10$  mpc and below $a=a_1=0.1$ mpc.
Setting $N=50$ and $m=50\msun$, the enclosed, distributed mass becomes
\beq
M_\star(<r) \approx 250\msun \tilde r,\ \ \ \ \tilde r \lap 10
\eeq
where $\tilde r$ is the radius in units of mpc and
the subscript ``$\star$'' indicates the distributed mass,
i.e., the stars.

While  the values of $N$ and $m/\mh$ were chosen primarily
on the basis of computational convenience, the models
adopted here are not necessarily poor representations of
real galactic nuclei.
Steady-state models of the center of the Milky Way
galaxy \cite{HA-06a,Freitag-06} typically find that the 
distributed mass within $\sim 100$ mpc of the 
MBH is dominated by stellar BHs
(as opposed other types of stellar remnant, or stars)
with $M_\star(r<10 \mathrm{mpc})\approx 10^3\msun$.
Expressed in terms of the gravitational radius defined
in Eq.~(\ref{eq:defrg}), and assuming $\rho\propto r^{-2}$,
a distributed mass of $10^3\msun$ within $10$ mpc implies
\beq
M_\star(<r) \approx 200 \left(\frac{r}{10^4r_g}\right).
\eeq
By comparison, the scaling adopted above implies that for
our models,
\beq
M_\star(<r) \approx 120 \left(\frac{r}{10^4r_g}\right).
\eeq

The $N$-body integrator is described elsewhere \citep{MM-06,MM-08};
it includes post-Newtonian accelerations of orders 
up to and including 2.5 (i.e. $c^{-5}$) in the interactions between the 
MBH- and star particles.
The algorithm was modified for the current study to allow
merger of star particles with the MBH.
The condition for a merger was an instantaneous separation 
\cite{Cutler-94}
\beq\label{eq:defrm}
r\le \rm=8r_g
\eeq
or $\sim 4\times 10^{-4}$ mpc if $\mh=10^6\msun$.
The angular momentum and eccentricity of an orbit that just
grazes the capture sphere are (in the Keplerian approximation)
\beq\label{eq:defem}
\lm = \sqrt{2G\mh\rm}, \ \ \ \ \eccm = 1-\frac{\rm}{a}.
\eeq
For $a_1\le a\le a_2$ and using the adopted value of $\rm$,
the eccentricity of a capture orbit satisfies
\beq
4\times 10^{-5} \lap 1-\eccm \lap 4\times 10^{-3}.
\eeq
The mass of a merged particle was added to that of the MBH
in such a way that linear momentum was conserved.

An EMRI event was defined as any merger occurring from an 
orbit with semi-major axis, at the moment of capture,
less than $0.01$ mpc.
Mergers occurring from orbits with larger $a$ were recorded
as ``plunges''.

While capture in its final stages would be driven by energy
loss due to emission of gravitational waves, as represented
here by the 2.5PN terms, the capture {\it rate} is determined primarily
by dynamical interactions that take place far beyond $r_g$.
In order to better understand the dynamical mechanisms leading
to capture, three series of $N$-body integrations were carried out, 
incorporating different subsets of the full PN equations of motion 
defined in Paper I.

\bigskip
\noindent
{\it Series I:} No PN terms were included. 
Stars were nevertheless allowed 
to merge with the MBH if they passed within $\rm$
(``plunges'').

\bigskip
\noindent
{\it Series II:} The 2.5PN terms were included.
As a result, some stars (``EMRIs'') were captured onto 
orbits for which the timescale for gravitational wave energy
loss is less than the timescale for scattering by other stars.

\bigskip
\noindent
{\it Series III:} All PN terms (1PN, 2PN, 2.5PN) were included.
\smallskip

\noindent
In each series, at least eight different Monte Carlo realizations of 
the same initial
conditions were integrated forward in time.
Models from Series I and II were integrated for a time of $10^7$ yr,
based on the scalings adopted above.
For models from Series III, inclusion of the additional PN terms
caused the integrator to run more slowly, and most integrations were
terminated after $2-3$ Myr.
Calculations were carried out on {\tt gravitySimulator}, the 32-node
cluster at RIT.

\section{Basic scales of length and time}

Here we define length and time scales associated with an idealized
model consisting of a central MBH and a smooth, spherical distribution
of surrounding stars.
Other time scales, associated with collisional (relaxation) processes, 
are defined below.

The length scale associated with the event horizon of the MBH is
\beq
r_g \equiv {G\mh\over c^2} \approx 4.80\times 10^{-5} \mathrm {mpc},
\label{eq:defrg}
\eeq
where the numerical value assumes $\mh=1.0\times 10^6\msun$.

Ignoring the contribution of the stellar BHs to the gravitational potential,
the (Newtonian) orbital period of a test mass around the MBH is
\beq\label{eq:period}
P_r = {2\pi a^{3/2}\over \sqrt{G\mh}}
\approx 2.96\ {\tilde a}^{3/2} \mathrm{yr}
\eeq
where $\tilde a$ is the test mass's semi-major axis in units 
of mpc;
the second relation again assumes $\mh=1.0\times 10^6\msun$.

Approximating the stellar BHs as a smooth mass distribution,
$\rho(r)=mn(r)$ with $n(r)$ given by Eq.~(\ref{eq:nofr}), 
their contribution to the gravitational potential is
\beq
\Phi_\star(r) = \frac{GM_0}{r_0}\ln\left(\frac{r}{r_0}\right) + \mathrm{constant}
\eeq
where $M_\star(<r)=M_0(r/r_0)$;
setting $r_0=1$ mpc gives $M_0=250\msun$ in our models.

Deviation of the potential from that of a point mass
induces a precession in the (fixed) plane of an orbiting star,
the ``mass precession.''
Orbital perturbation theory \cite[e.g.][]{MV-10} gives for
the precession rate, in the limit $M_0\ll\mh$,
\beq\label{eq:nuM}
\frac{d\omega}{dt} \equiv \nu_\mathrm{M} \approx 
-\nu_r \frac{M_\star(r<a)}{\mh}
\frac{\sqrt{1-e^2}}{1+\sqrt{1-e^2}}
\eeq
where
\begin{eqnarray}
\nu_r &\equiv& \frac{2\pi}{P_r} = (G\mh)^{1/2} a^{-3/2} \nonumber \\
&\approx& (0.47 \mathrm{yr})^{-1} {\tilde a}^{-3/2}
\end{eqnarray}
is the radial frequency and $e$ is the orbital eccentricity.
Precession is retrograde, i.e. in the opposite sense to the orbital motion.
For the adopted mass model \footnote{Throughout this paper we use $P$ to denote a full period
of oscillation or libration, while $t_X\equiv\pi/\nu_x$ is the time
for angular variable $X$ to change by $\pi$.},
\begin{subequations}
\begin{eqnarray}
\label{eq:nup}
t_\mathrm{M} &\equiv& \left|\frac{\pi}{\nu_\mathrm{M}}\right|
\approx \left(1.18\times 10^4\mathrm{yr}\right) g(e) {\tilde a}^{1/2} , \\
\label{eq:defgofe}
g(e) &=& \frac{1+\sqrt{1-e^2}}{2\sqrt{1-e^2}}.
\end{eqnarray}
\end{subequations}
In the limit $e\rightarrow 1$, Eq.~(\ref{eq:nuM}) predicts
$\nu_\mathrm{M}\rightarrow 0$, i.e. radial orbits do not precess.

The post-Newtonian accelerations also contribute to the
in-plane precession.
To lowest order, the Schwarzschild contribution is
\beq\label{eq:nuGR}
\nu_\mathrm{GR} = 
\frac{3}{c^2}\frac{\left(G\mh\right)^{3/2}}{\left(1-e^2\right)a^{5/2}}
= \nu_r \frac{3 r_g}{a(1-e^2)}\eeq
in the opposite (prograde) sense, and
\beq
t_\mathrm{GR}\equiv\left|\frac{\pi}{\nu_{\mathrm{GR}}}\right|
\approx \left(1.02\times 10^4 \mathrm{yr}\right)\
\left(1-e^2\right) \tilde a^{5/2}.
\eeq
While we defer a detailed treatment of spin effects to a subsequent paper,
we note here that the Kerr contribution to the in-plane precession is smaller
than (\ref{eq:nuGR}) by a factor $\sim \chi(r_g/p)^{1/2}$,
where $\chi$ is the dimensionless spin parameter of the MBH and 
 $p=(1-e^2)a$ is the semi-latus rectum.
Excepting very shortly before a merger, this factor would be
much smaller than unity in our simulations.


\section{Series I and II}
\label{sec:seronetwo}

\begin{figure*}
\includegraphics[width=12.cm]{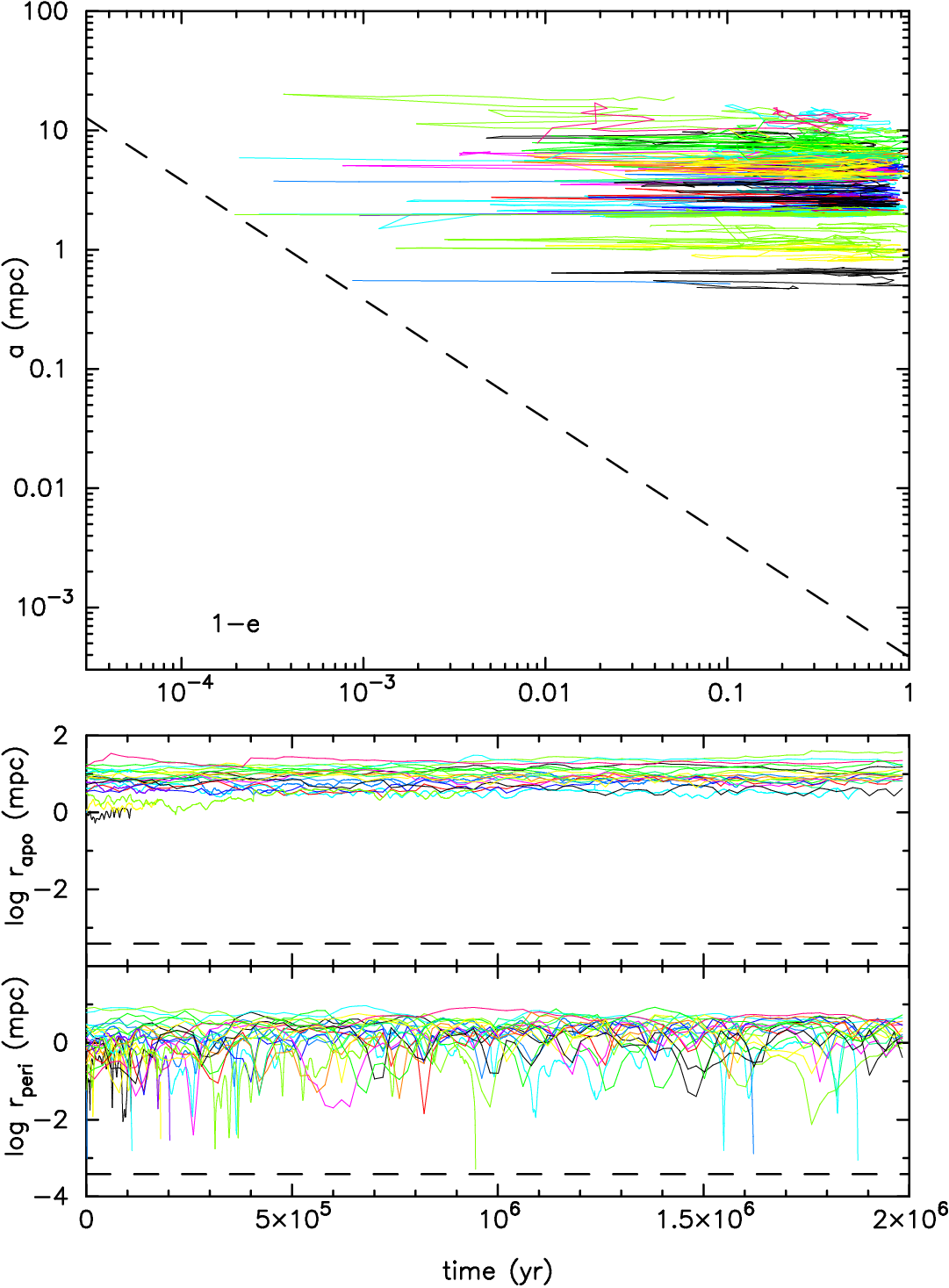}
\caption{\label{fig:ser1} 
A simulation from Series I (Newtonian).
$a$ and $e$ are the semi-major axis and
eccentricity respectively of the two-body system consisting
of one star and the MBH;
$r_\mathrm{peri}=(1-e)a$ and $r_\mathrm{apo}=(1+e)a$.
Different colors correspond to different particles
(since the total number of colors available was 12, 
each color is used for more than one particle).
Dashed lines indicate the capture radius, $\rm=8r_g$.
}
\end{figure*}

As discussed in more detail below, including the
relativistic terms in the $N$-body equations of motion
resulted in much lower EMRI rates than expected based
on Newtonian dynamics of a compact cluster around a MBH.
The essential element that differs between the relativistic
and non-relativistic dynamics turns out to be 
the 1PN precession of the periapse,
Eq.~(\ref{eq:nuGR}).
In order to quantify the magnitude of the differences,
two sets of experiments
were carried out in which some or all of the relativistic
terms were omitted.
Integrations from Series I were based on the Newtonian 
equations of motion.
Series II included also the 2.5PN terms, allowing 
capture of stars onto inspiral orbits via
GW energy loss.
In integrations from both Series I and Series II, 
the 1PN (Schwarzschild) precession is absent.

\subsection{Series I}

Fig.~\ref{fig:ser1} shows, for one integration in Series I,
the evolution of semi-major axis $a$ and eccentricity $e$
for each of the 50 stars, until a time of $2$ Myr.
Star-star gravitational scattering induces substantial
changes in the orbital angular momenta
over these time scales, while the energy (i.e.
semi-major axis) remains nearly constant.

Whenever the periapse distance $r_\mathrm{p}\equiv a(1-e)$ 
drops below $\rm$ the star is captured.
Almost all such events are ``plunges'' since no GW energy loss occurs, 
and since essentially no stars have initial semi-major axes
less than $0.01$ mpc, the condition defined above for a
capture to be classified as an EMRI.
In the simulation shown in Fig.~\ref{fig:ser1},
17 out of the 50 stars are captured by $t=2$ Myr
 and 30 stars are captured by $t=10$ Myr.

\begin{figure*}
\includegraphics[width=6.cm,angle=-90.]{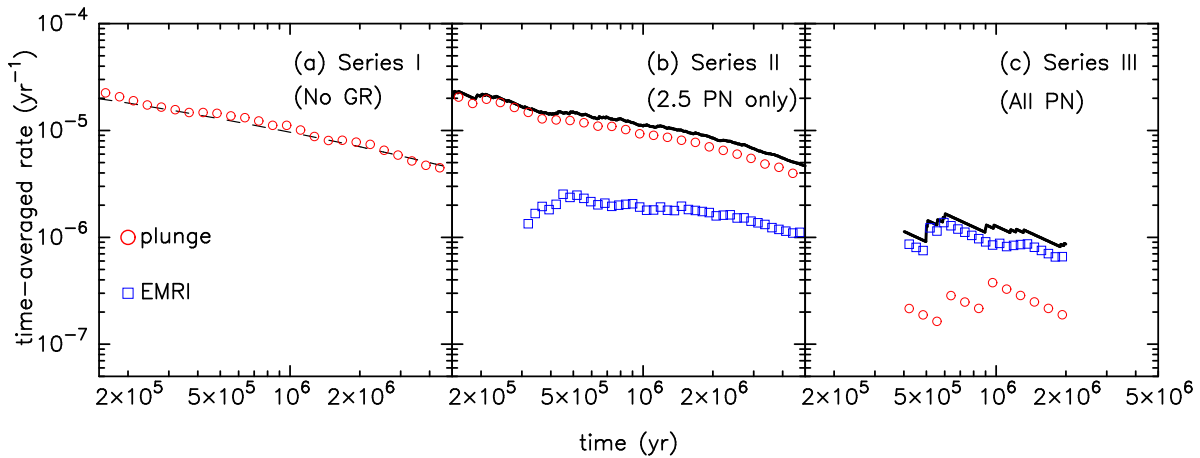}
\caption{\label{fig:rates} 
Time-averaged capture rates, defined as the total number of
events until time $t$ divided by $t$, computed from the complete 
set of runs in each series.
{\it Series I:} dashed line is the prediction of 
Eqs.~(\ref{eq:dNdtau})-(\ref{eq:nsol}) for $C_1=0.5$.
The solid (black) curves in the other panels are the
total event rates, EMRI plus plunges.
}
\end{figure*}

Fig.~\ref{fig:rates} shows the time-averaged 
capture rate as a function of time,
defined as the number of mergers occurring in time $t$
divided by $t$.
Events from each simulation in the series were summed and
the result was divided by the number of simulations; 
in other words, the plotted rates refer to a cluster
with $N(t=0)=50$.
The capture rate drops with time, since both the number of stars
available to merge, and the number of stars acting as
scatterers, decrease with time.

In this (Newtonian) regime, the mechanism expected to 
dominate the scattering of stars onto high-eccentricity
orbits around a point mass is resonant relaxation (RR) 
\cite{Rauch-96,HA-06b}.
Because the smooth gravitational potential has symmetries
that restrict the orbital evolution (i.e. to fixed ellipses
in the case of a Newtonian point mass),
perturbations on a test star are not random but correlated.
This leads to coherent changes $\Delta\mathbf{L}=\mathbf{T}t$
on times $t\lap \tc$ by the residual torque 
$\left|\mathbf{T}\right|\approx\sqrt{N}Gm/r$ exerted by
the $N$ randomly oriented, orbit-averaged mass distributions
of the surrounding stars.
The coherence time is set by the mechanism that most
rapidly causes orbits to precess, randomizing $\mathbf{T}$.
In these non-relativistic simulations, that mechanism
is mass precession, Eq.~(\ref{eq:nup}).
The accumulated change over $\tc$, 
$\left|\Delta\lc\right|\sim\left|\mathbf{T}\tc\right|$, 
then becomes the mean free path in $\mathbf{L}$ space for the 
long-term ($t\gg\tc$) random walk in $\mathbf{L}$.
The effective relaxation time associated with RR satisfies
\beq
\frac{\left|\Delta\lc\right|}{L_c}\sqrt{\frac{t}{\tc}}
\equiv \sqrt{\frac{t}{\trr}}
\eeq
i.e.
\beq\label{eq:deftrr}
\trr = \left(\frac{L_c}{\Delta\lc}\right)^{2}\tc 
\eeq
where $L_c\equiv \sqrt{GM_\bullet a}$ is the angular momentum of a 
circular orbit with radius $r\approx a$.
These relations should be understood as correct only in an
order-of-magnitude sense.

In the coherent regime the change in orbital angular momentum is
\begin{equation}\label{eq:dLdt}
\left|\frac{dL}{dt}\right| \approx
\sqrt{N}\,\frac{Gm}{a} 
\approx \beta_s \frac{L_c}{P}\,\frac{\sqrt{N}\,m}{M_\bullet} 
\end{equation} 
where  
$N$ is roughly the number of stars 
within a sphere of radius $a$,
and $\beta_s$ is a dimensionless factor of order unity
\cite{GH-07,Eilon-09}.

The precession time due to the distributed mass,
Eq.~(\ref{eq:nup}), can be written
\beq\label{eq:deftM}
\tm \approx (1.2\times 10^4 \mathrm{yr}) g(e) {\tilde a}^{1/2}
\left(\frac{\mh}{10^6\msun}\right) 
\left(\frac{M_0}{250\msun}\right)^{-1} 
\eeq
where $M_\star(<r)=M_0\tilde r$ and $M_0=250\msun$ 
for the models considered here.

Identifying $\tm$ with $\tc$, and writing $N(<r)=M_\star(<r)/m$,
Eqs.~(\ref{eq:deftrr})-(\ref{eq:deftM}) give
\begin{subequations}\label{eq:trr}
\begin{eqnarray}\label{eq:trra}
&&\trr \approx \beta_s^{-2} g(e)^{-1} \frac{\mh}{m} P(a) 
\\\label{eq:trrb}
&&\approx \frac{5.9\times 10^4 \mathrm{yr}}{\beta_s^2 g(e)}
\left(\frac{\mh}{10^6\msun}\right)^{1/2}
\left(\frac{m}{50\msun}\right)^{-1}
{\tilde a}^{3/2}
\end{eqnarray}
\end{subequations}
We note that $M_\star$ has dropped out.
This is only valid for values of $M_\star$ large enough
that $\tc=t_\mathrm{M}$ is shorter than all other time scales of interest.

In the expression for $\trr$, the form of the density profile is 
still reflected in the
dependence of $g$ on $e$, which, for the assumed initial
mass distribution, is strongly dependent on $e$
as $e\rightarrow 1$ (Eq.~\ref{eq:defgofe}).
However what matters for the coherence breaking is the 
relative precession of the test particle's orbit with
respect to the other orbits.
Hence it is reasonable to average $g(e)$ in Eq.~(\ref{eq:trr})
over the eccentricity distribution for all the stars, 
Eq.~(\ref{eq:Nofae}):
\beq
\overline{g} \equiv \int_0^1 g(e) de^2 = \frac32.
\eeq
In what follows we ignore changes in $\overline{g}$ due to evolution of
$M_\star(r)$.

Because the integration time is long compared with $\trr$,
we expect a quasi-steady-state to be set up in the angular momentum
distribution at each $a$, such that the rate of loss of stars
into the capture sphere is roughly equal to the rate at which
new stars are being scattered onto low-angular-momentum orbits.
In this regime, the (differential) rate at which stars are scattered into
the MBH at each $a$ is given approximately by:
\beq\label{eq:Gamma}
\Gamma(a,t) da \approx \frac{N(a,t)da}{\ln\left(L_c/L_\mathrm{m}\right)\trr}
\eeq
\cite[e.g.][]{CK-78}.
The logarithmic term can be interpreted as the
approximate number of relaxation times required for an
orbit to diffuse in angular momentum from $e\approx 0$ to
$e\approx \eccm$ \cite{LS-77}.

\begin{figure}
\includegraphics[width=8.5cm]{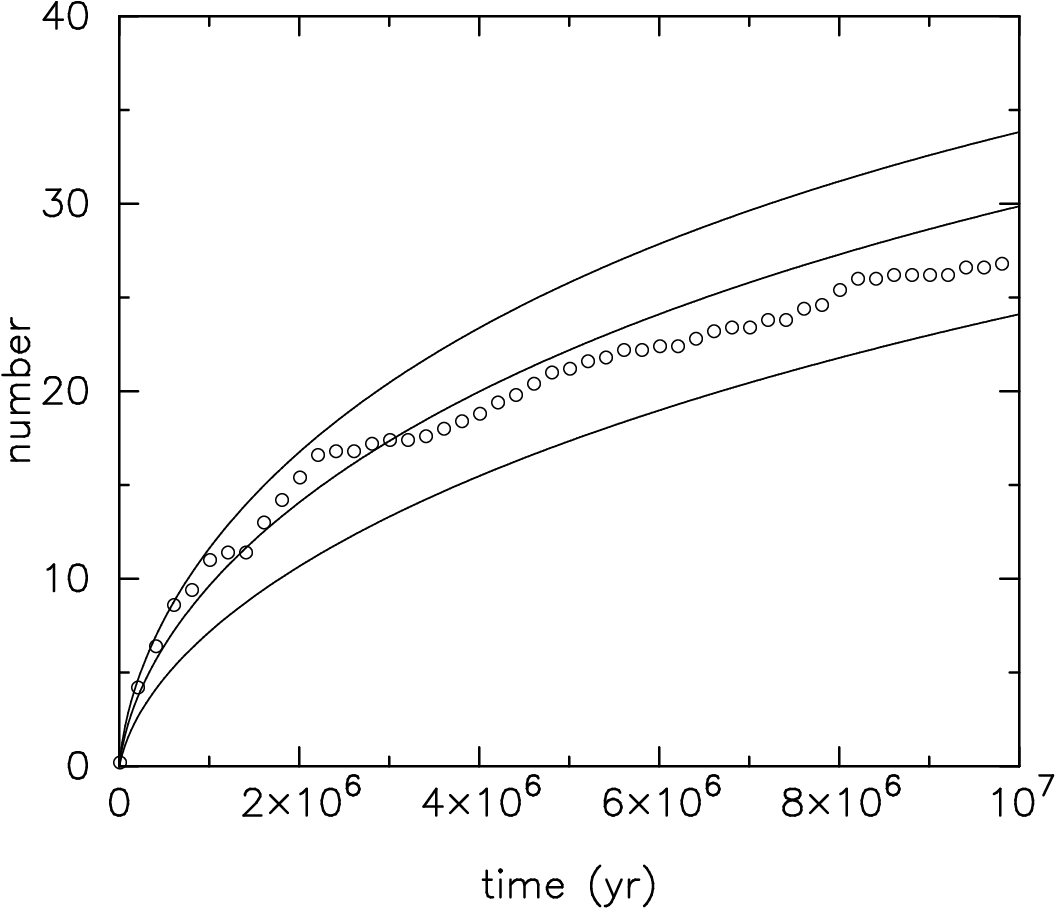}
\caption{\label{fig:rr} 
Open circles show the number of stars captured up until time $t$ 
in the combined runs from Series I; values are normalized to an initial
total number of 50.
Plotted events are all ``plunges.''
Solid lines show the predictions of Eqs.~(\ref{eq:dNdtau})-(\ref{eq:nsol}) for $C_1=(0.3,0.5,0.7)$.
}
\end{figure}

Using Eq.~(\ref{eq:trr}), the differential capture rate 
can be written
\beq
\Gamma(a,t) da = C_1 \frac{m}{\mh} \frac{N(a,t)da}
{\ln\left(1-e_\mathrm{m}^2\right)^{-1/2}P(a)}
\eeq
where $\beta_s^2\overline{g}$ and all other
uncertainties have been absorbed into the fitting parameter $C_1$,
assumed independent of $a$ and $t$.

Equating $\Gamma(a,t)$ with $-dN(a,t)/dt$ and using 
Eqs.~(\ref{eq:defem})
and~(\ref{eq:period}),
the evolution equation for $N(a,t)$ becomes
\begin{subequations}
\label{eq:dNdtau}
\begin{eqnarray}
&&\frac{\partial N}{\partial\tau} = 
-\frac{N(a,\tau)}{{\tilde a}^{3/2} \ln\left(a/\rm\right)}, 
\ \ \ \ \tau\equiv t/t_1, \\
t_1 &=& (5.9\times 10^4 \mathrm{yr})\, C_1^{-1}
\frac{\mh/m}{2\times 10^4}
\left(\frac{\mh}{10^6\msun}\right)^{-1/2}.
\end{eqnarray}
\end{subequations}
with solution
\begin{subequations}
\label{eq:nsol}
\begin{eqnarray}
N(\tilde a,\tau) &=& N(\tilde a,0) e^{-\tau/\tau_1}, \\
\tau_1 &=& {\tilde a}^{3/2} \ln\left(a/\rm\right).
\end{eqnarray}
\end{subequations}
Fig.~\ref{fig:rr} plots the predicted, cumulative 
number of events versus time, compared with the results 
from the Series I integrations.
The agreement is good for $C_1\approx 0.5$.
The mean capture rate is given by 
$t^{-1}\int_{a_1}^{a_2}\left[N(a,0)-N(a,t)\right]da$;
this is plotted, with $C_1=0.5$, as the dashed line
in Fig.~\ref{fig:rates}a.

\begin{figure}
\includegraphics[width=8.5cm]{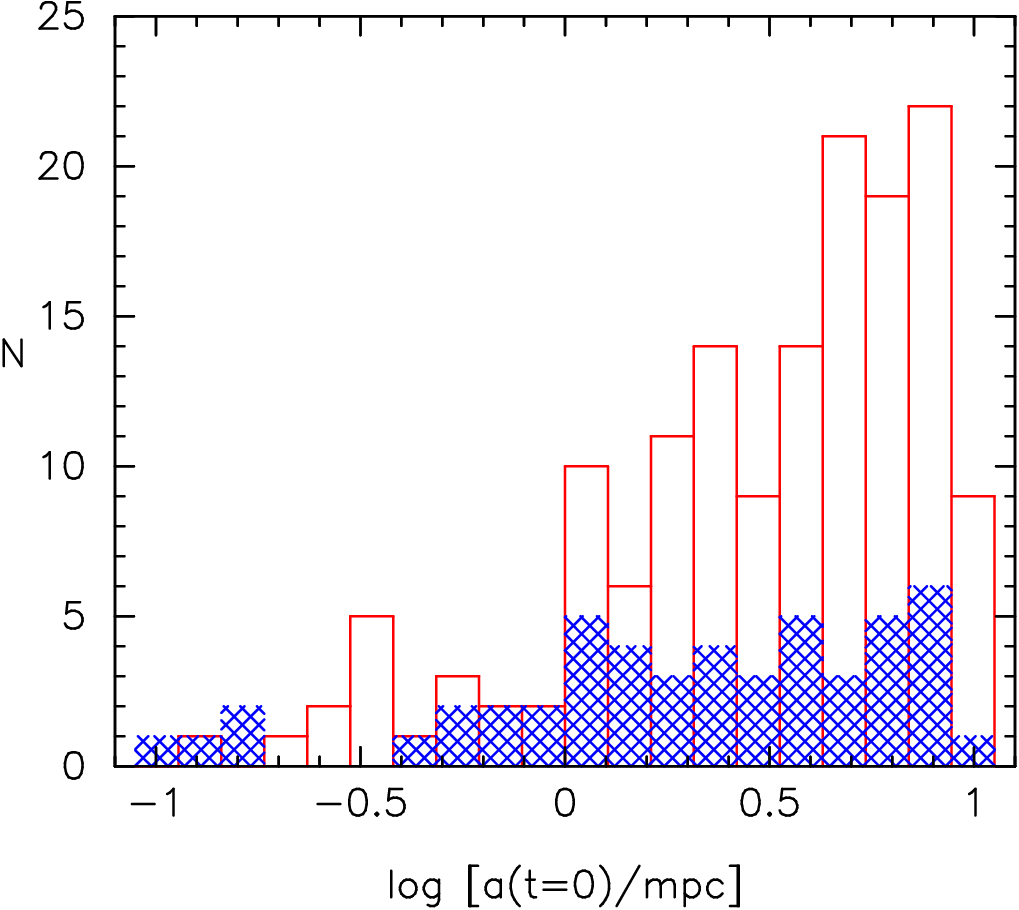}
\caption{\label{fig:hists} 
Distribution of initial semi-major axes for the 
capture events from Series II.
Red (unfilled) histogram shows the plunges;
blue (cross-hatched) histogram shows the EMRIs.
Elapsed time is $10^7$ yr.}
\end{figure}

\subsection{Series II}

Series II integrations included the 2.5 PN terms in the equations
of motion,
allowing some stars to be captured onto orbits that inspiral gradually into
the MBH via GW energy loss.

The energy loss timescale associated with the 2.5PN terms is \cite{Peters-64}
\label{eq:deftgw}
\begin{eqnarray}
t_\mathrm{GW} &\equiv& \left|\frac{1}{a}\frac{da}{dt}\right| \\
&=&\frac{5}{64} \frac{c^5a^4}{G^3\mh^2m}\left(1-e^2\right)^{7/2}
\left(1+\frac{73}{24}e^2 + \frac{37}{96}e^4\right)^{-1}\nonumber \\
&\approx& 1.2\times 10^{14}\mathrm{yr} 
\left(\frac{m}{50\msun}\right)^{-1}
\left(\frac{\mh}{10^6\msun}\right)^{-2}
{\tilde a}^4 \left(1-e\right)^{7/2} \nonumber
\end{eqnarray}
where the latter expression assumes $e\approx 1$.
In this limit, GW inspiral occurs along lines of fixed slope 
in the $a, (1-e)$ plane until shortly before the merger:
\beq\label{eq:deda}
\frac{\Delta(1-e)}{1-e} \approx -\frac{\Delta a}{a},
\eeq
such that $r_p=(1-e)a$ is approximately constant \citep{Peters-64}.
In order to avoid plunging, a star must reach a high enough
eccentricity that the GW timescale is shorter
than the time for gravitational encounters to scatter the star onto 
a different orbit.

From Eqs.~(\ref{eq:deftgw})-(\ref{eq:deda}),
the time required for GWs to change $e$ by of order $1-e$ is
$\sim t_\mathrm{GW}$.
In the case of gravitational encounters, changes in angular momentum
are equivalent to changes in eccentricity since $a$ is nearly conserved.
The time $t_L$ for encounters to change $L$ by of order itself is
\beq
(\Delta L)^2 \approx L^2 = L_c^2 \frac{t_L}{\trr}
\eeq
i.e.
\beq\label{eq:tl}
t_L = \left(\frac{L}{L_c}\right)^2\trr \approx 2(1-e)\trr
\eeq
with $\trr$ the RR timescale defined above.
Equating $t_L$ with $t_\mathrm{GW}$ then gives the condition for
capture onto an  inspiral orbit:
\beq\label{eq:cond}
a\left(1-e\right) \approx \frac{1}{2}\left(\frac{340\pi}{3}\right)^{2/5} r_g
\eeq
i.e. capture requires a periapse distance of $\sim 5r_g$.
This is slightly smaller than the separation at which
mergers were assumed to take place in the simulations (Eq.~\ref{eq:defrm}).

Given the approximate nature of Eq.~(\ref{eq:cond}), 
one expects capture onto inspiral orbits
for some fraction, of order unity,
of stars that would otherwise plunge into the MBH.
Fig.~\ref{fig:hists} shows the cumulative histogram
of capture events for the integrations of Series II.
Roughly $1/4$ (54 out of 206) events were EMRIs.
Time-averaged capture rates are shown in Fig.~\ref{fig:rates}b.
These results are consistent with expectations.

As shown in the next section, these results are substantially changed
by the inclusion of the 1PN and 2PN terms into the equations of motion.

\section {\label{sec:serthree}Series III}

\begin{figure*}
\includegraphics[width=12.cm]{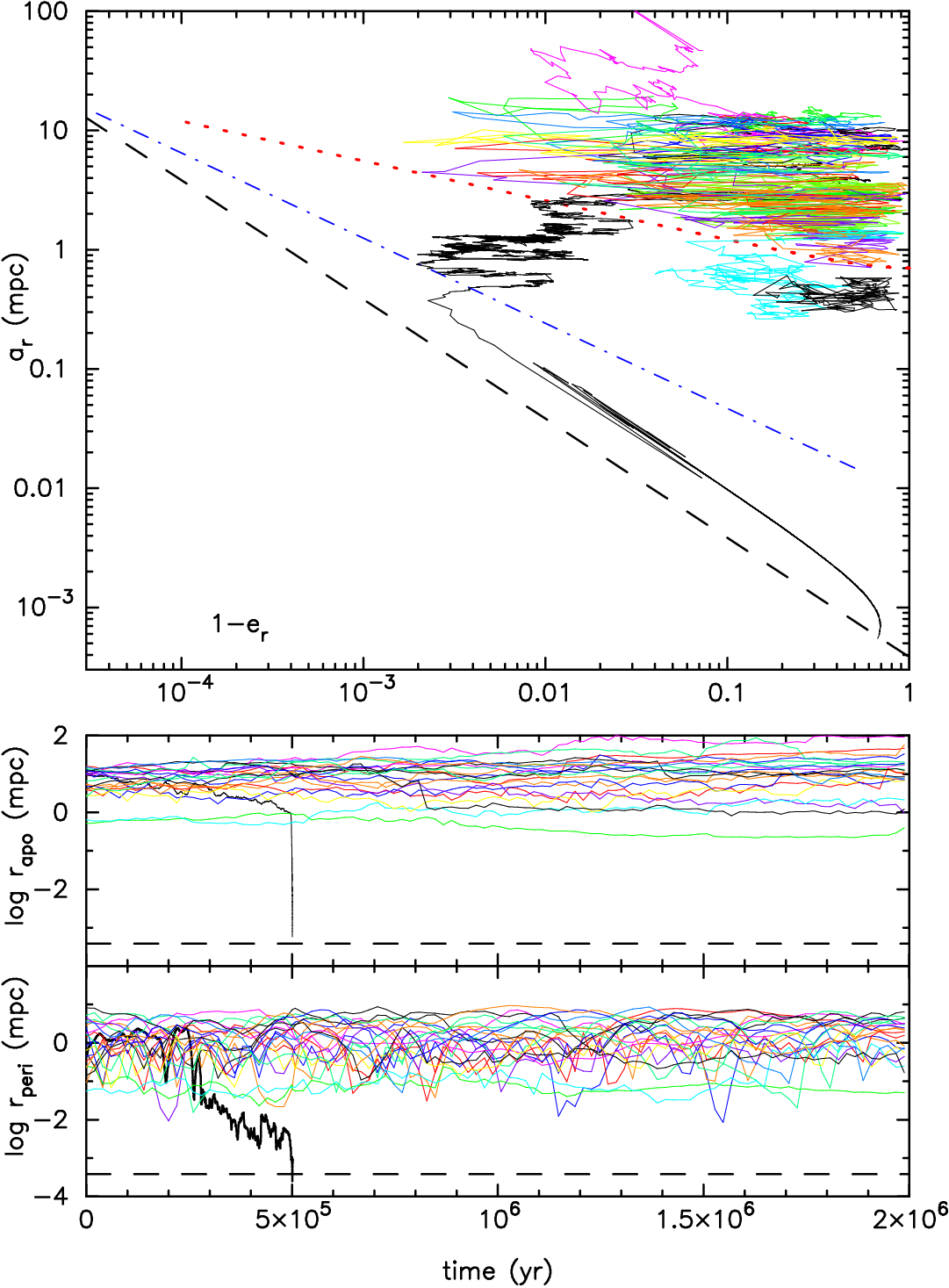}
\caption{\label{fig:ser3} 
A simulation from Series III (all PN terms).
$a_r$ and $e_r$ are the 1PN generalizations of the semi-major axis and
eccentricity;
$r_\mathrm{peri}=(1-e_r)a_r$ and $r_\mathrm{apo}=(1+e_r)a_r$.
Different colors correspond to different particles
(the number of different colors is 12 so  
each color is used for more than one particle).
Dashed (black) lines show the assumed capture radius, $\rm=8r_g$.
In  the top frame, the dotted (red) line is the Schwarzschild barrier,
Eq.~(\ref{eq:SB}), and the dash-dotted (blue) line is the 
approximate condition for GW capture, Eq.~(\ref{eq:cond2}).}
\end{figure*}

Fig.~\ref{fig:ser3} shows the evolution on the 
($a_r, 1-e_r$) plane of all 50 stars in an integration from Series III.
Integrations in this series included all PN terms (1PN, 2PN, 2.5PN).
The quantities $a_r, e_r$ are the 1PN generalizations of the
Keplerian semi-major axis and eccentricity respectively; to this PN order,
the periapse and apoapse distances are given respectively by
$(1-e_r)a_r$ and $(1+e_r)a_r$
\citep{DD-85}.

Over the course of the 2 Myr interval plotted in Fig.~\ref{fig:ser3}, 
only one capture occurs: an EMRI.
The plunge events that dominated the integrations from Series I
and II are absent.
The mean capture rates computed from all integrations in 
Series III are shown in Fig.~\ref{fig:rates}.
The mean capture rate is $\lap 1$ Myr$^{-1}$, with 74\% of the events EMRIs.
By comparison, in Series I and II, the mean capture rate was
$\gap 10$ Myr$^{-1}$ at early times,
and almost all events were plunges.

\begin{figure*}
\includegraphics[width=7.cm,angle=-90.]{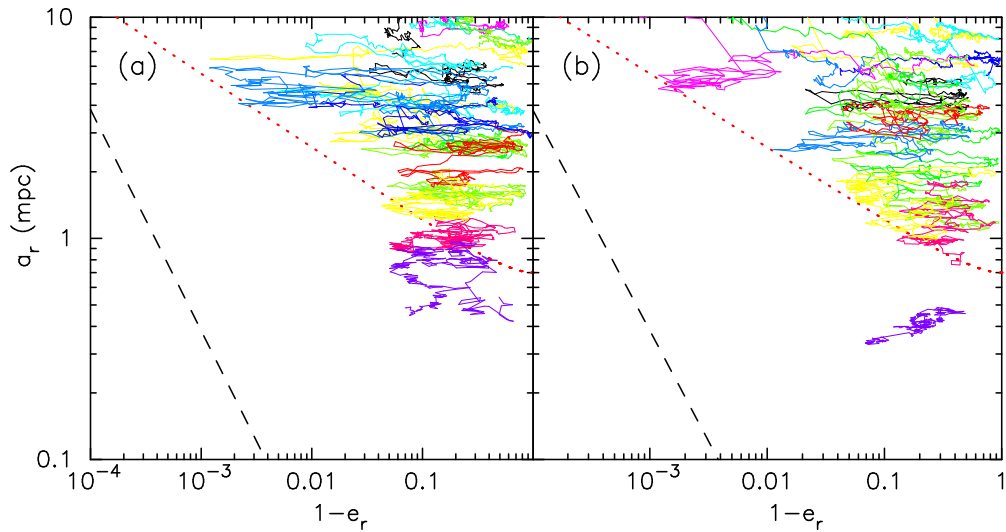}
\caption{\label{fig:2parts} 
Illustrating the angular momentum barrier when all PN terms are included,
in two short time segments extracted from a Series III  integration.
(a) $4\times 10^5 \mathrm{yr} \le t \le 8\times 10^5 $ yr;
(b) $1\times 10^6$ yr $\le t \le 1.4\times 10^6$ yr.
Dashed (black) line is the capture radius;
dotted (red) line is the predicted angular momentum barrier, Eq.~(\ref{eq:SB}).
Stars that lay initially to the left of the barrier were excluded.
This integration produced no EMRIs.
The time interval plotted, $\Delta t = 4\times 10^5$ yr,
is somewhat longer than the RR timescale of Eq.~(\ref{eq:trr})
and much longer than $(1-e_r)\trr$.
Note the ``barrier penetration'' at small values of $a_r$.
}
\end{figure*}

The proximate reason for the much lower event rate in 
the integrations from Series III
is suggested by Fig.~\ref{fig:ser3}:
there is an apparent barrier in orbital eccentricity,
or angular momentum, which very few stars cross on 
Myr timescales.
Furthermore, the single star that is captured in that figure
appears to require a time much longer
than $\sim (1-e_r)\trr$ to cross the gap.
The barrier is illustrated more clearly in Fig.~\ref{fig:2parts},
based on another integration from Series III.
This figure shows that there is some ``barrier penetration''
for orbits with small semi-major axis; we discuss the reasons
below.


\subsection{The Schwarzschild barrier}

Adding the 1PN terms to the equations of motion results
in precession of the argument $\omega$ of orbital periapse, 
with an orbit-averaged frequency given by Eq.~(\ref{eq:nuGR}).
For low eccentricity orbits, 
the rate of this Schwarzschild precession \footnote{So-called to distinguish it from precession associated with frame-dragging, or Kerr precession, in which the 
line of nodes changes.
Other names for precession induced by the Schwarzschild part of the metric
include geodetic and de Sitter precession.}
is comparable to
that produced by the distributed mass, Eq.~(\ref{eq:nuM}),
at the radii of interest here.
But whereas the latter rate tends to zero as $e\rightarrow 1$,
the Schwarzschild precession rate diverges, as $\sim (1-e)^{-1}$.
The effective time over which background torques can act 
is determined by the fastest 
mechanism that changes the relative orientation of a star 
with respect to the gravitational field produced by all the other stars.
For a highly eccentric orbit, this mechanism is Schwarzschild precession
and its associated timescale tends to zero as $e\rightarrow 1$.

We suggest that the angular momentum barrier be identified,
in a qualitative way,
with the value of $L$ at which the torques
become ineffective due to the orbit's rapid Schwarzschild precession.

\begin{figure}
\includegraphics[angle=-90.,width=9.cm]{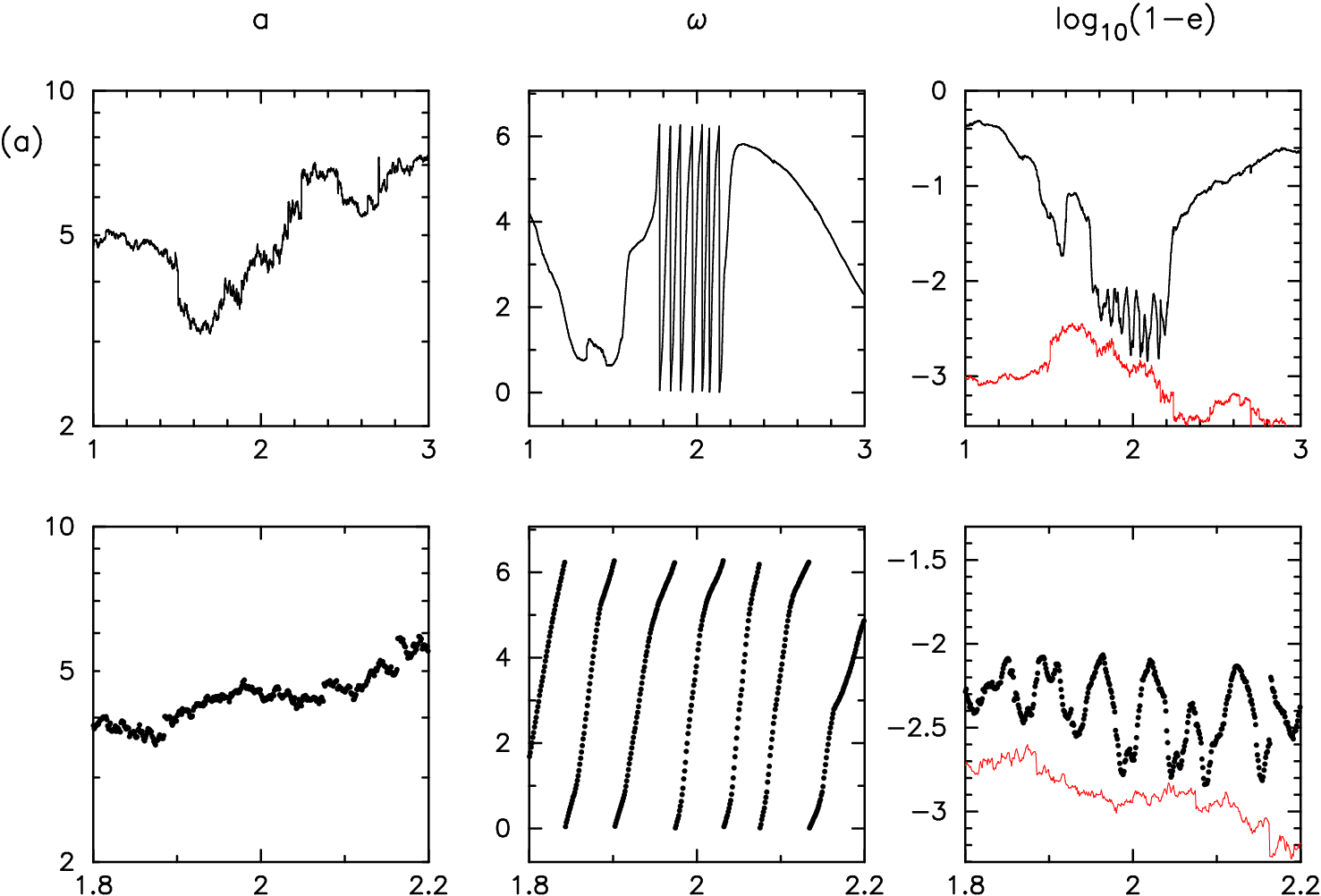}
\vskip 0.15truein
\includegraphics[angle=-90.,width=9.cm]{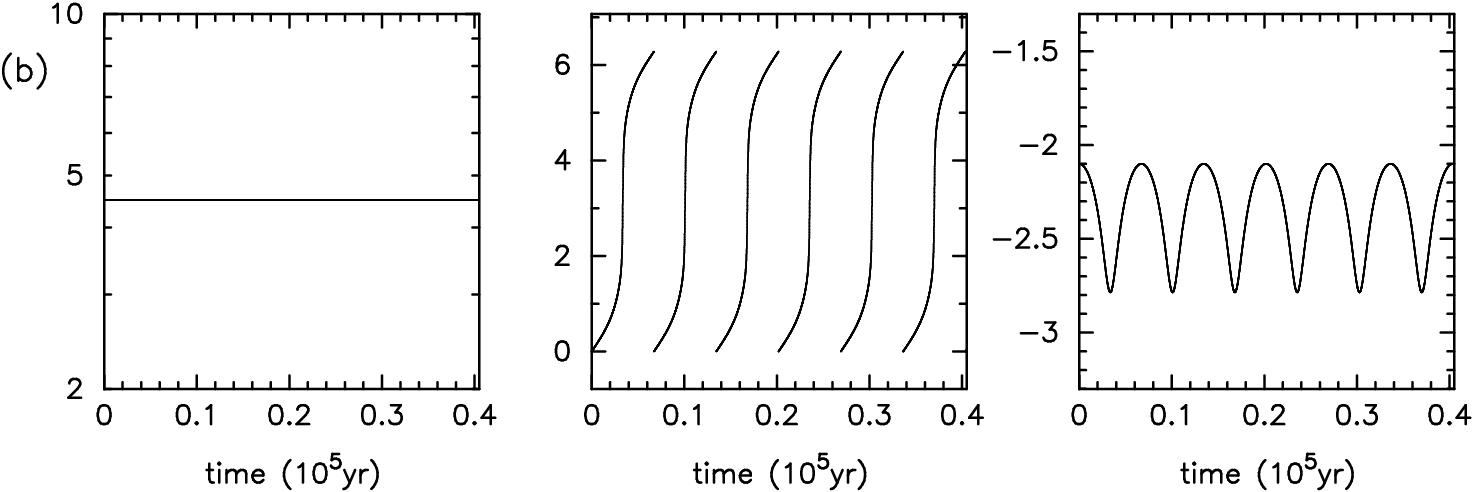}
\caption{\label{fig:Nbodybounce} 
(a) A ``bounce'' orbit extracted from a Series III $N$-body integration.
Plotted are the semi-major axis, argument of periapse and eccentricity versus time,
at low (upper) and high (middle) time resolutions.
In the plots of eccentricity vs. time, the lower (red) curves
show the predicted location of the angular momentum  barrier, 
Eq.~(\ref{eq:SB}), with $C_\mathrm{SB}=0.7$.
Changes in the predicted barrier location reflect changes in the
semi-major axis.
(b) A solution to the equations of motion 
(\ref{eq:motion}) that
reproduces the important features of the $N$-body orbit in (a).
The duration of the ``bounce'' phase is roughly the coherence
time for the background (stellar) potential.
Additional details are given in the text.}
\end{figure}

The residual torque produced by an otherwise-spherical distribution
of stars, at $r\approx a$, is of order
\beq\label{eq:RRtorque}
T\approx \frac{Gm}{a} \sqrt{N(a)}
\approx \frac{1}{\sqrt{N(a)}}\frac{GM_\star(a)}{a}
\eeq
where $M_\star(a)=mN(a)$ is the distributed mass within radius $r=a$.
Writing $L=\left[G\mh a(1-e^2)\right]^{1/2}$ for the angular
momentum of a test star, 
the time scale over which this fixed torque changes $L$ is
\beq
\left|\frac{1}{L}\frac{dL}{dt}\right|^{-1} \approx 
\sqrt{N(a)}\frac{\mh}{M(a)}
\left[\frac{a^3(1-e^2)}{G\mh}\right]^{1/2}.
\eeq
The condition that this time be shorter than the relativistic
precession time, $\pi/\nu_\mathrm{GR}$, is 
\beq\label{eq:barrier}
\ell> \ell_\mathrm{SB}\approx 
\frac{r_g}{a} \frac{\mh}{M_\star(a)}\sqrt{N(a)}
\eeq
where we have written $\ell\equiv L/L_c=(1-e^2)^{1/2}$.
Evaluating the quantities in Eq.~(\ref{eq:barrier})
for the $N$-body models, the critical semi-major axis 
becomes
\beq\label{eq:SB}
{\tilde a} = C_\mathrm{SB} \left(1-e^2\right)^{-1/3}
\eeq
where $C_\mathrm{SB}$ is a constant of order unity.
Eq.~(\ref{eq:SB}), with $C_\mathrm{SB}=0.7$, 
is plotted as the dotted (red)
lines in Figs.~\ref{fig:ser3}, \ref{fig:2parts}, \ref{fig:Nbodybounce}
and~\ref{fig:gw}.

Assuming that the condition~(\ref{eq:SB}) holds for all
values of $a$, the normalizing constant $C_\mathrm{SB}$
can be interpreted as the value of $\tilde a$ when $e=0$,
i.e. as the minimum value of $\tilde a$ for which the barrier exists.
One expects that orbits with semi-major axes larger than this
minimum value (and smaller than a maximum value, defined below),
and that approach the barrier from the right
on Figs.~\ref{fig:ser3} or~\ref{fig:2parts},
will have a hard time crossing it,
since torques become inefficient near the barrier.
 
We develop these ideas more quantitatively in the next subsection.
Before doing so we present a more quantitative model for
the behavior of low-angular-momentum orbits under the combined 
influence of relativistic precession and Newtonian torques.

Fig.~\ref{fig:Nbodybounce}a shows the evolutionary track
of a star from a Series III integration.
The star first strikes the barrier at $t\approx 1.8\times 10^5$ yr;
the eccentricity then oscillates several times at roughly
fixed amplitude before decreasing again, carrying the
star away from the barrier at $t\gap 2.2\times 10^5$ yr.
During each bounce, the argument of periapse $\omega$ advances
by $\sim 2\pi$.

Many other examples of ``bounce'' near the angular momentum barrier
were extracted from the $N$-body integrations.
While differing in detail, all such orbits exhibited a variation in
eccentricity near the barrier with a period roughly equal
to the period of Schwarzschild precession.

This feature suggests that the torques responsible for 
angular momentum changes near the barrier are due to a 
distortion of the stellar potential, expressed about the
location of the MBH particle, that is lop-sided or dipole
in character. 
(A quadrupole distortion would cause changes in $L$ 
at twice the frequency of circulation, etc.).
That the dominant component of the torquing potential
should have such a form is not unreasonable, 
since if one represents the gravitational potential 
from $N$ orbit-averaged stars in a multipole expansion, 
the largest terms are expected be the monopole (due to the
spherical cluster) followed by the dipole (due to $\sqrt{N}$ 
departures from spherical symmetry) etc.

We also verified that the behavior of orbits like that plotted
in Fig.~\ref{fig:Nbodybounce} was unchanged if the mass of the
test particle was drastically reduced.
These tests confirmed that the variations in orbital angular
momentum in the test particle's orbit were not a spurious 
result of motion of the black hole
particle induced by the test star's precession.

We used the following simple potential to model the motion
of a test star subject to a lopsided force from all the other stars:
\begin{equation}
\Phi({\boldsymbol r}) = -\frac{G\mh}{r} +
\Phi_s V(r) - S(a) a \cos\theta.
\label{eq:Potential}
\end{equation}
The second term on the right hand side of~(\ref{eq:Potential})
is the potential of the spherical star cluster.
For the models considered here,
\beq
V(r) = \ln\left(\frac{r}{r_0}\right), \ \ \Phi_s = \frac{GM_0}{r_0}, \ \ M_0=M_\star(r<r_0).
\eeq
The third term represents the lopsided distortion of the stellar
potential; the amplitude $S$, which has dimensions of acceleration,
 is assumed to depend on the test-orbit's
semi-major axis as 
\begin{equation}\label{eq:Sofa}
S(a)\approx \frac{Gm\sqrt{N(a)}}{a^2}\approx \frac{GM_\star(a)}{a^2\sqrt{N(a)}}.
\end{equation}
The corresponding density is
\beq
\rho_D(r,\theta) = \frac{S(a)a}{2\pi G} \frac{\cos\theta}{r^2}.
\eeq
The integrated mass corresponding to the lopsided component is zero.

Due to the dominance of the first term in Eq.~(\ref{eq:Potential}), 
the radial period of an orbit is shorter than all other orbital time scales.
In Appendix~\ref{appendB} we express the Hamiltonian 
corresponding to the potential~(\ref{eq:Potential}) in terms of Delaunay
action-angle variables and average over the the
radial motion, including the orbit-averaged term that generates
the Schwarzschild precession.
The result is a set of four equations that describe the rates
of change of the (osculating) Keplerian elements ($L,L_z,\omega,\Omega$):
\begin{subequations}
\label{eq:motion}
\begin{eqnarray}
\label{eq:motiona}
\frac{d\omega}{d\tau} &=& \ell^{-2} -A_M\ell/(1+\ell) + \nonumber \\
&&A_D\sin\omega\left[
-\frac{\ell}{e}\sin i + \frac{e}{\sin i}\frac{\ell_z^2}{\ell^3}\right], \\
\frac{d\ell}{d\tau} &=& -A_D e \sin i \cos\omega, 
\label{eq:motionb}
\\
\frac{d\Omega}{d\tau} &=&  
-A_D e \frac{\ell_z}{\ell^2} \frac{\sin\omega}{\sin i}, 
\label{eq:motionc}
\\
\frac{d\ell_z}{d\tau} &=& 0 \label{eq:motiond}
\end{eqnarray}
\end{subequations}
where we have defined the dimensionless elements
$\ell = L/I = \sqrt{1-e^2}$,
$\ell_z = L_z/I$ and $\cos i = \ell_z/\ell$
in terms of the radial action $I=(G\mh a)^{1/2}$,
and the dimensionless time is $\tau\equiv\nu_0t$, where
\beq
\nu_0 = \nu_r \frac{3G\mh}{c^2a}.
\eeq

In defining the Delaunay variables, the plane of reference 
has been taken to be the ($x,y$) plane and the reference direction
is the $x$-axis; thus an orbit in the ($x,z$) plane has
$\sin i = 1$, and for such an orbit, $\omega=\pi/2$
corresponds to an orientation parallel to the $z$-axis,
the assumed direction of the lopsided distortion.
In the case of an orbit in the ($x,y$) plane, $\sin i=0$,
$\ell_z=\ell$, and the orientation of the orbit is determined by
$\omega+\Omega$; according to Eqs.~(\ref{eq:motiona}) and
(\ref{eq:motionc}), the terms in $(d/d\tau)(\omega+\Omega)$ 
that are proportional to $A_D$ sum to zero in this case.

The dimensionless parameters $A_M$ and $A_D$ specify the
strength of the spherical and lopsided components of the
distributed mass:
\begin{subequations}
\label{eq:defAmAd}
\begin{eqnarray}
A_M &=& \frac{1}{3}\frac{M_\star(a)}{\mh} \frac{a}{r_g},
\label{eq:defAmAda}
\\ 
A_D &=& \frac{1}{3} \frac{S}{G\mh/a^2} \frac{a}{r_g} 
\approx \frac{1}{3\sqrt{N}} \frac{M_\star(a)}{\mh} 
\frac{a}{r_g}.\label{eq:defAmAdb}
\end{eqnarray}
\end{subequations}
We note that $A_M/A_D \approx N^{1/2}$ which is of order
unity in the models considered here.
Thus whenever it is relevant to neglect $A_M$, $A_D$ is also negligible.

After averaging, the first and third terms in Eq.~(\ref{eq:Potential}) 
result in the same equations of motion as in the classical Stark problem 
\citep[e.g.][]{LL-76}.
The corresponding solutions \citep[e.g.][]{BelyaevRafikov-10} 
consist of circulation in $\Omega$, with period
\beq
P_\mathrm{Stark} = \frac{4\pi}{3S}\sqrt{\frac{G\mh}{a}};
\eeq
oscillations  in $i$, $e$ and $\omega$ have the same period, while
the $z$-component of the angular momentum is fixed.
The eccentricity reaches a maximum value that depends on $L_z$;
for $L_z=0$, i.e. for $i=\pi/2$, $e_\mathrm{max}=1$,
while for $L_z\ne 0$ the maximum eccentricity is less than one.

In the case considered here, precession of a sufficiently eccentric orbit 
is dominated by the
Schwarzschild term, the first term on the right hand side of
Eq.~(\ref{eq:motiona}).
Such an orbit circulates in a nearly fixed plane 
and the eccentricity varies with a period equal to the period of circulation.
If $\ell$ is sufficiently small, 
we can write $d\omega/d\tau\approx \langle\ell\rangle^{-2}$ 
with $\langle\ell\rangle$ the time-averaged angular momentum.
Eqs.~(\ref{eq:motion}) then have approximate solution
\begin{subequations}
\label{eq:approxell}
\begin{eqnarray}
1 - \frac{\ell(t)}{\langle\ell\rangle} &\approx& \langle\ell\rangle
A_\mathrm{D}\sin i\cos\left(\nu t\right), \\
\nu &=& \frac{3}{c^2}\frac{\left(G\mh\right)^{3/2}}{\langle\ell^2\rangle a^{5/2}}.
\end{eqnarray}
\end{subequations}
In this limit, the amplitude of the angular momentum oscillations,
\beq\label{eq:l-l+}
\ell_+ - \ell_- \approx 2\langle\ell\rangle^2 A_\mathrm{D}\sin i ,
\eeq
decreases quadratically with $\langle\ell\rangle$.

We now return to  the full equations of motion~(\ref{eq:motion})
in order to test whether the detailed behavior of stellar
trajectories near bounce in the $N$-body integrations is consistent with
our simple model.
In the $N$-body models, the dimensionless parameters that appear 
in the equations of motion are
\begin{subequations}
\label{eq:params}
\begin{eqnarray}
A_M &\approx& 1.8 {\tilde a}^2, \label{eq:paramsa}\\
A_D &\approx& 1.2 {\tilde a}^{3/2}, \label{eq:paramsb}\\
\nu_0 &\approx& (3.26\times 10^3 \mathrm{yr})^{-1} {\tilde a}^{-5/2}
\label{eq:paramsc}
\end{eqnarray}
\end{subequations}
and the Schwarzschild precession period is
\beq\label{eq:tGRNbody}
P_\mathrm{GR}\equiv \frac{2\pi}{\nu_\mathrm{GR}} 
\approx 2.1\times 10^4\mathrm{yr}\
(1-e^2){\tilde a}^{5/2}.
\eeq
The $N$-body orbit in Fig.~\ref{fig:Nbodybounce}a 
exhibits $\sim 6$ full circulations in
$\omega$ in a time $\sim 3.5\times 10^4$ yr, 
corresponding to a precessional period of 
$\sim 6\times 10^3$ yr.
The semi-major axis for this star during the bounce is
$3.5\lap\tilde a\lap 5$ and the eccentricity is
$-2.8\lap \log_{10}(1-e)\lap -2.1$.
Inserting $\tilde a=4$ and $ \log_{10}(1-e)=-2.5$ into
Eq.~(\ref{eq:tGRNbody}) gives 
$P_\mathrm{GR}\approx 4\times 10^3$ yr which is quite consistent
with the observed precessional period.

Fig.~\ref{fig:Nbodybounce}b shows a solution to Eqs.~(\ref{eq:motion})
that reproduces the other important features of the $N$-body orbit 
near bounce.
We set $A_\mathrm{D}=10$ and $A_\mathrm{D}=30$ 
i.e., $\tilde a\approx 4$; 
the initial values of the orbital elements were
$\log(1-e)=-2.1$, $\omega=-\pi/2$, $\Omega=0$, and $i=0.35\pi$.
Variations in $\Omega$ and $i$ (not shown here)
were similar in amplitude for 
the $N$-body and numerically computed orbits.

We carried out similar comparisons for other orbits near the barrier.
While differing in details, all the cases examined could be adequately
represented via solutions to our simple Hamiltonian~(\ref{eq:hamil}).

\begin{figure}
\includegraphics[width=8.cm]{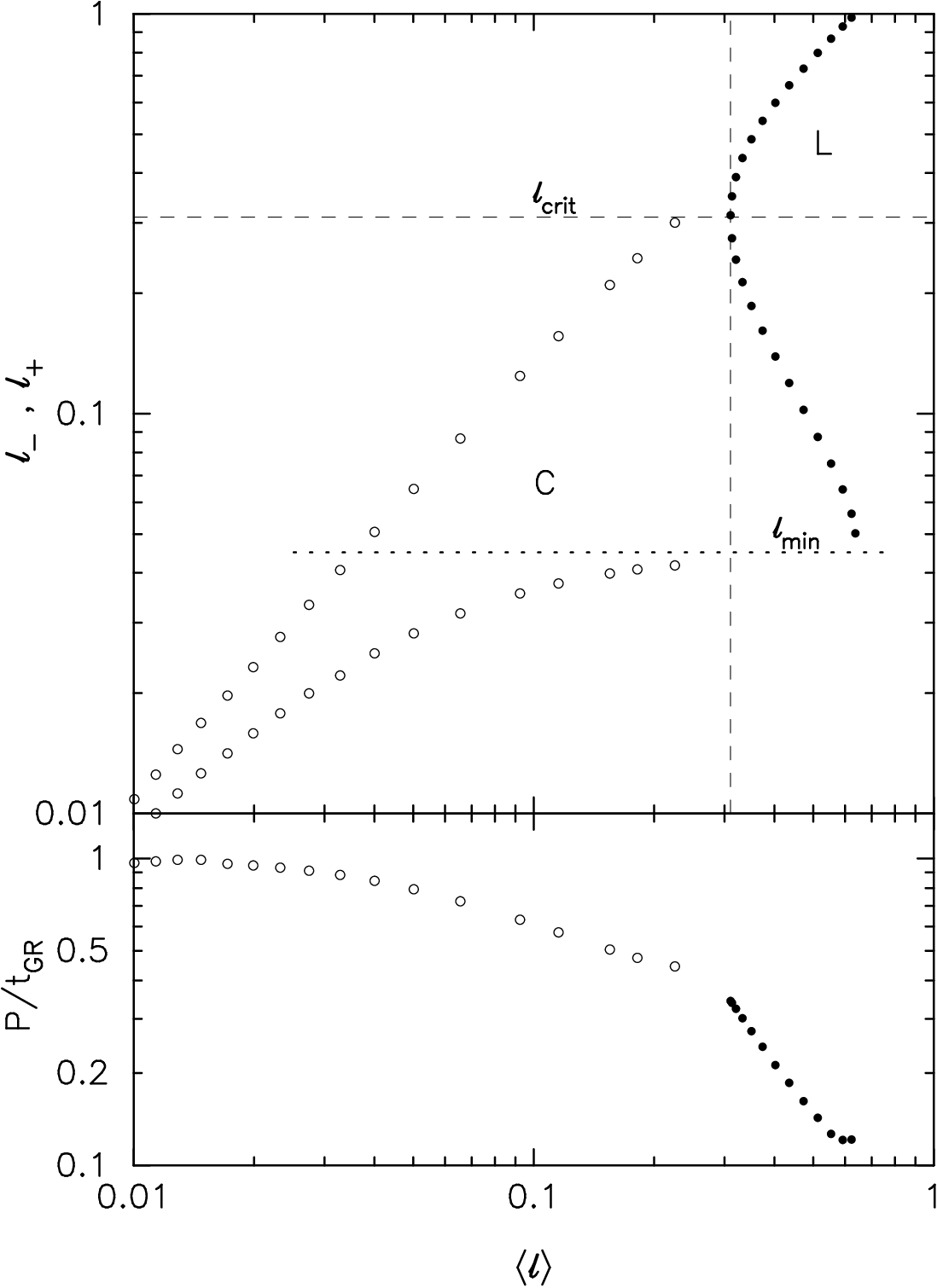}
\caption{\label{fig:2dorbits} 
Properties of two-dimensional ($\sin i = \pi/2$) solutions
to the equations of motion~(\ref{eq:motion}),
with $A_\mathrm{M}=30, A_\mathrm{D}=10$.
$\langle\ell\rangle$ is the dimensionless angular momentum,
$\ell=\sqrt{1-e^2}$, 
averaged over one precessional or librational period $P$.
{\it Top panel:} minimum and maximum angular momenta reached
by the orbit over one period.
Filled circles (``L'') are librating orbits while open circles
(``C'') are circulating orbits.
The dashed lines at $\ell=\ell_\mathrm{crit}$ mark the angular momentum
at which the precession rate, $d\omega/dt$, is zero.
The dotted line marked $\ell_\mathrm{min}$ is the minimum 
angular momentum reached by librating orbits; it is argued
in the text that this is essentially the angular momentum corresponding
to the Schwarzschild barrier, Eq.~(\ref{eq:barrier}).
{\it Bottom panel:} Librational/precessional periods as a 
fraction of the Schwarzschild period, 
computed from Eq.~\ref{eq:nuGR} after replacing $(1-e^2)$ by 
$\langle\ell\rangle^2$. }
\end{figure}

\begin{figure}
\includegraphics[width=8.5cm]{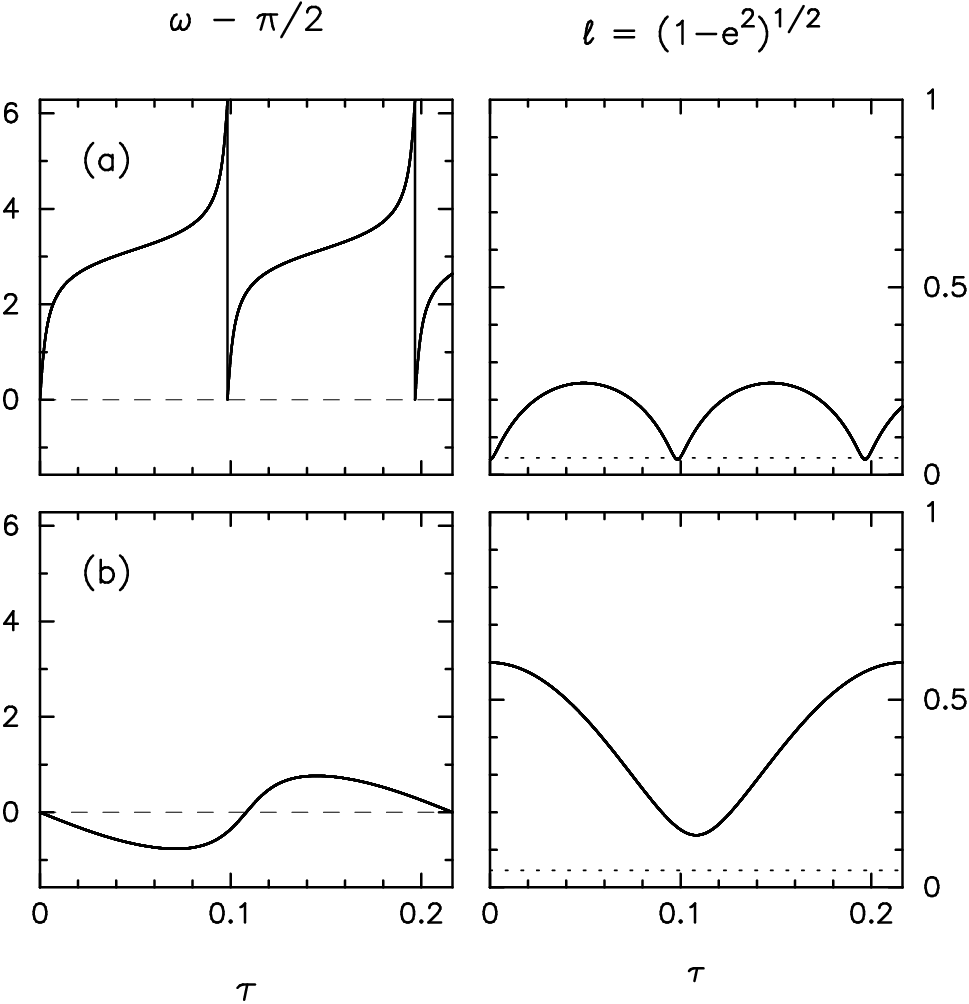}
\caption{\label{fig:orbits} 
Two orbits from Fig.~\ref{fig:2dorbits}. 
(a) $\langle\ell\rangle = 0.225$ (circulating); 
(b) $\langle\ell\rangle = 0.402$ (librating).
$\omega=\pi/2$ corresponds to orientation along the $z$ axis. 
The dotted lines in the right-hand panels indicate $\ell_\mathrm{min}$.}
\end{figure}

The assumptions made in deriving the Hamiltonian~(\ref{eq:hamil})
are not specific to orbits near the barrier.
Fig.~\ref{fig:2dorbits} summarizes the properties of orbits, of
arbitrary angular momentum but restricted to the $x-z$ plane
($\sin i = \pi/2$),
in the potential of Fig.~\ref{fig:Nbodybounce}b.
Orbits can either circulate (small angular momentum) or librate
(large angular momentum).
There is a critical value of the angular momentum, at the time
the orbit is oriented parallel to the $z$-axis (i.e. the direction
of the lopsided distortion),
such that the precession rate $\dot\omega=0$ and the orbit
remains fixed in orientation; from Eq.~(\ref{eq:motiona}) this 
occurs when $\ell=\ell_\mathrm{crit}$ where
\beq
0 = 1 - A_\mathrm{M}\frac{\ell_\mathrm{crit}^3}{1+\ell_\mathrm{crit}} - 
A_\mathrm{D}\frac{\ell_\mathrm{crit}^3}{\sqrt{1-\ell_\mathrm{crit}^2}}
\eeq
or $\ell_\mathrm{crit}\approx 0.310$ in Fig.~\ref{fig:2dorbits}.
Away from this value, 
librating orbits experience both their minimum and maximum
angular momenta when precessing past the $z$-axis; first in
one sense (when the mass precession term dominates)
and then in the other (when the Schwarzschild term dominates).
Libration changes to circulation  when the orbit precesses
by an angle $\pm\pi$ from its starting point along the $z$-axis.
The minimum angular momentum reached by the orbit at the
libration/circulation boundary 
is labelled $\ell_\mathrm{min}$ on Fig.~\ref{fig:2dorbits};
in this potential, $\ell_\mathrm{min}\approx 0.045$.

Fig.~\ref{fig:orbits} shows two orbits from this plane,
one circulating and one librating.
We note here one property common to both types of orbit:
stars tend to spend more time with high angular momentum than
with low angular momentum.
The reasons for this behavior are apparent in the equations of 
motion~(\ref{eq:motion}).
(1) For large $\ell$, $d\ell/d\tau$ is small, i.e. the 
orbit-averaged effects of the torque are small.
(2) Orbits tend to linger at 
values of $\omega$ corresponding to large $\ell$
since the Schwarzschild precession rate is proportional
to $\ell^{-2}$.
(The latter trend reverses for orbits so circular that
mass precession dominates the Schwarzschild precession.)

Combining Eqs.~(\ref{eq:barrier}) and (\ref{eq:defAmAdb}),
\beq\label{eq:lSB}
\ell_\mathrm{SB}\approx (2A_\mathrm{D})^{-1}
\eeq
or $\ell_\mathrm{SB}\approx0.05$ for 
$A_\mathrm{D}=10$.
Not coincidentally, this is roughly equal to $\ell_\mathrm{min}\approx 0.045$.
Fig.~\ref{fig:2dorbits} shows that $\ell_\mathrm{min}$ specifies
not only the minimum angular momentum achievable by librating orbits,
but is also roughly the minimum $\ell$ reached by orbits 
whose angular momentum changes by of order itself over
one period; these were the two assumptions made in deriving
(\ref{eq:barrier}).

At the same time, it is clear from Fig.~\ref{fig:2dorbits} 
that orbits with $\ell_-\ll \ell_\mathrm{min}$ do exist.
Apparently, such orbits are rarely reached in the $N$-body
integrations.
We discuss the reasons in the next sub-section.

\subsection{Barrier penetration. I.}

Here we address the question of how orbits evolve
after striking the angular momentum barrier.
The mechanism (``tunneling'') explored in this section, 
which is based on resonant relaxation,
will turn out to be less important as a source of barrier
penetration than the mechanism presented in the next section, 
based on non-resonant relaxation.
We nevertheless explore it in some detail since doing so
will lead to insights about why the barrier is so ``hard''
on time scales comparable to the RR time.

The Hamiltonian model just presented assumed a fixed gravitational potential.
The (constant) term responsible for the torques was assumed to
arise from the time-averaged potential of the $N$ stars.
In reality, the background potential must change as the orbits
of all the stars evolve (e.g. precess), leading to quasi-random changes 
in the direction and amplitude of the torque that acts on a single star.
The changes in the background torques are responsible both for
moving a star toward the barrier and moving it away.
As we show here, such changes can also result in barrier penetration.

In Sec.~\ref{sec:seronetwo}, the mass precession time,
\begin{subequations}
\begin{eqnarray}
\label{eq:tM2}
\tm &\equiv&\frac{\pi}{\left|\nu_\mathrm{M}\right|} \nonumber \\
&\approx& 1.8\times 10^4 \mathrm{yr}\ {\tilde a}^{1/2}
\left(\frac{\mh}{10^6\msun}\right) 
\left(\frac{M_0}{250\msun}\right)^{-1},
\end{eqnarray}
\end{subequations}
was taken as the ``coherence time'' over which the background
torques can be assumed constant.
We emphasize that the relevant time here is the precession 
time for a typical orbit, hence we have set $g(e)=\overline g = 3/2$
in Eq.~(\ref{eq:deftM}); the fact that some (high-$e$)
orbits precess much faster due to relativity is not important.

A second relevant time scale is 
the so-called vector resonant relaxation (VRR) time,
the time for
orbital planes to change due to their mutual torques
\citep[e.g.][]{HA-06a}:
\begin{eqnarray}
\label{eq:deftRRv}
t_\mathrm{RR,v} &\approx& \frac{\pi}{\nu_r} \frac{\mh}{m\sqrt{N}} 
\nonumber \\
&\approx& 1.3\times 10^4 \mathrm{yr}\ {\tilde a}
\left(\frac{\mh}{10^6\msun}\right)^{-1} 
\left(\frac{\tilde N}{5}\right)^{-1/2}
\end{eqnarray}
where we have written $\tilde N = N(<\tilde a)/\tilde a$;
$\tilde N \approx 5$ in the $N$-body models. 
Since $t_\mathrm{M}\approx t_\mathrm{RR,v}/\sqrt{N}$,
one normally assumes $t_\mathrm{M}\ll t_\mathrm{RR,v}$.
However the small values of $N$ considered here, together
with the approximate nature of Eq.~(\ref{eq:deftRRv}),
 means that the two time scales are essentially the same 
in these models at all radii of interest.
Furthermore, VRR leads to full randomization of orbital
orientations in the sense that it changes orbital planes
as well as orientations within the plane;
mass precession leaves the orbital planes unchanged.
For these reasons, we adopt 
$t_\mathrm{RR,v}$ as the coherence time in the remainder
of this section.

The time-independent model presented above implicitly assumed that
$t_\mathrm{coh}$ was long compared with orbital precessional periods.
In fact, the ratio of the coherence time to the 
Schwarzschild precession time is
\begin{subequations}
\begin{eqnarray}
\frac{t_\mathrm{coh}}{t_\mathrm{GR}} &\approx&
3\frac{r_g}{a}\frac{\mh}{m}\frac{1}{\sqrt{N}}\frac{1}{1-e^2}\\
&\approx& \frac{3}{2} A_\mathrm{D}^{-1} \left(1-e^2\right)^{-1}.
\end{eqnarray}
\end{subequations}
Using Eq.~(\ref{eq:barrier}) to relate $a$ to $e$
along the barrier,
\begin{subequations}
\label{eq:tratio}
\begin{eqnarray}
\label{eq:tratioa}
\left(\frac{t_\mathrm{coh}}{t_\mathrm{GR}}\right)_\mathrm{SB} &\approx&
3\frac{a}{r_g}\frac{m}{\mh}\sqrt{N} \\
\label{eq:tratiob}
&\approx& 6 A_\mathrm{D} \approx 7.2 {\tilde a}^{3/2}
\end{eqnarray}
\end{subequations}
where the final expression refers to the $N$-body models
and uses Eq.~(\ref{eq:paramsb}).
Eqs.~(\ref{eq:tratio}) suggest that for orbits near the
barrier with $\tilde a\gtrsim 1$,
the background potential 
should remain constant for several Schwarzschild
precessional periods.
This is consistent with the observed behavior
of orbits like the one in Fig.~\ref{fig:Nbodybounce}a
($\tilde a\approx 4$),
which precesses a few times before (presumably) changes in
the background potential cause the orbit to evolve
away from the barrier.
In the case of orbits with $\tilde a\lesssim 1$
the two time scales can be assumed to be comparable
in our models.

One way to penetrate the barrier is suggested by
Fig.~\ref{fig:2dorbits}.
Random changes in the background potential, e.g. in the
direction of the lopsided term, could have the effect of moving
orbits progressively to the left and downward on that plot.
This is because the minimum angular momentum reached by
an orbit over a precessional period, $\ell_-$, depends
both on the instantaneous value of $\ell$, and on the relative orientation
of the orbit and the torquing term.
A sequence of correlated changes in the direction of the torque
could result in gradual transition down the narrow
``neck'' at the lower left of the diagram, toward
arbitrarily small values of $\langle\ell\rangle$.

This mechanism can be simply modeled if we  assume that
(1) changes in the background potential are instantaneous,
separated by time $\sim t_\mathrm{coh}$, and 
(2) $t_\mathrm{coh}\gg t_\mathrm{GR}$, so that
the orbital phase is essentially random at the time that
the potential changes.
The first assumption is not likely to be satisfied 
in all cases and we relax it below; however we will argue
that it corresponds to the highest probability for
barrier penetration.

Consider first the two-dimensional case, i.e. $\sin i = \pi/2$.
Assume as well that the orbit is sufficiently far down the
``neck'' that the angular momentum follows 
Eq.~(\ref{eq:approxell}),
\beq\label{eq:elapprox}
\ell^0(\omega) \approx \langle\ell\rangle^0
\left[1-\langle\ell\rangle^0  A_\mathrm{D}
\cos(\omega-\omega_\mathrm{D}^0)\right]
\eeq
where $\omega_D^0$ is the initial orientation of the 
lopsided distortion.
Now let the direction of the distortion instantaneously change,
 to $\omega_\mathrm{D}^1$.
The new orbit follows
\beq
\ell^1(\omega) \approx \langle\ell\rangle^1
\left[1-\langle\ell\rangle^1  A_\mathrm{D}
\cos(\omega-\omega_\mathrm{D}^1)\right].
\eeq
If the change occurs when $\omega=\omega^1$ then
\begin{eqnarray}
\langle\ell\rangle^1\left[1-\langle\ell\rangle^1 A_\mathrm{D}
\cos(\omega^1-\omega_\mathrm{D}^1)\right] &=& \nonumber \\
\langle\ell\rangle^0\left[1-\langle\ell\rangle^0 A_\mathrm{D}
\cos(\omega^0-\omega_\mathrm{D}^0)\right]&.&
\end{eqnarray}
The change in $\langle\ell\rangle$, $\Delta\langle\ell\rangle
\equiv\langle\ell\rangle^1 - \langle\ell\rangle^0$ is
\beq
\Delta\langle\ell\rangle \approx -\langle\ell\rangle^2 A_\mathrm{D}
\left[\cos\delta(1-\cos\Delta) - \sin\delta\sin\Delta\right]
\eeq
where $\delta = \omega^1-\omega_\mathrm{D}^0$ and 
$\Delta = \omega_\mathrm{D}^1 - \omega_\mathrm{D}^0$;
in this simple model, both angles are random variables.
Decreases in $\langle\ell\rangle$ are clearly allowed,
although the amplitude of the step size 
becomes increasingly small,
$\Delta\langle\ell\rangle\sim \langle\ell\rangle^2$, 
as $\langle\ell\rangle$ decreases.

There is an additional reason why evolution toward small
$\langle\ell\rangle$ is disfavored.
Eqs.~(\ref{eq:elapprox}) assume a constant rate of circulation in 
$\omega$ and ignore the eccentricity dependence of the 
averaged torque, Eq.~(\ref{eq:motionc}).
But as noted above, away from the limit of small $\langle\ell\rangle$,
orbits violate both assumptions, and
tend to linger at high values of $\ell$.
As a result, changes in the potential are most likely to occur
when $\ell$ is large.

For these two reasons, we do not expect the mechanism discussed in
this section to be
effective at moving stars very far to the left of the Schwarzschild
barrier; indeed we will argue in a subsequent paper 
\citep{AM-11} that evolution to lower
$\ell$ via this mechanism is exponentially suppressed
(and this is the basis for assigning the name ``tunneling''.) 
However, the ineffectiveness of RR at breaching the barrier is
important in explaining why the barrier is observed to be so ``hard,''
at least on time scales comparable to $t_\mathrm{RR}$.

\begin{figure}
\includegraphics[width=7.cm]{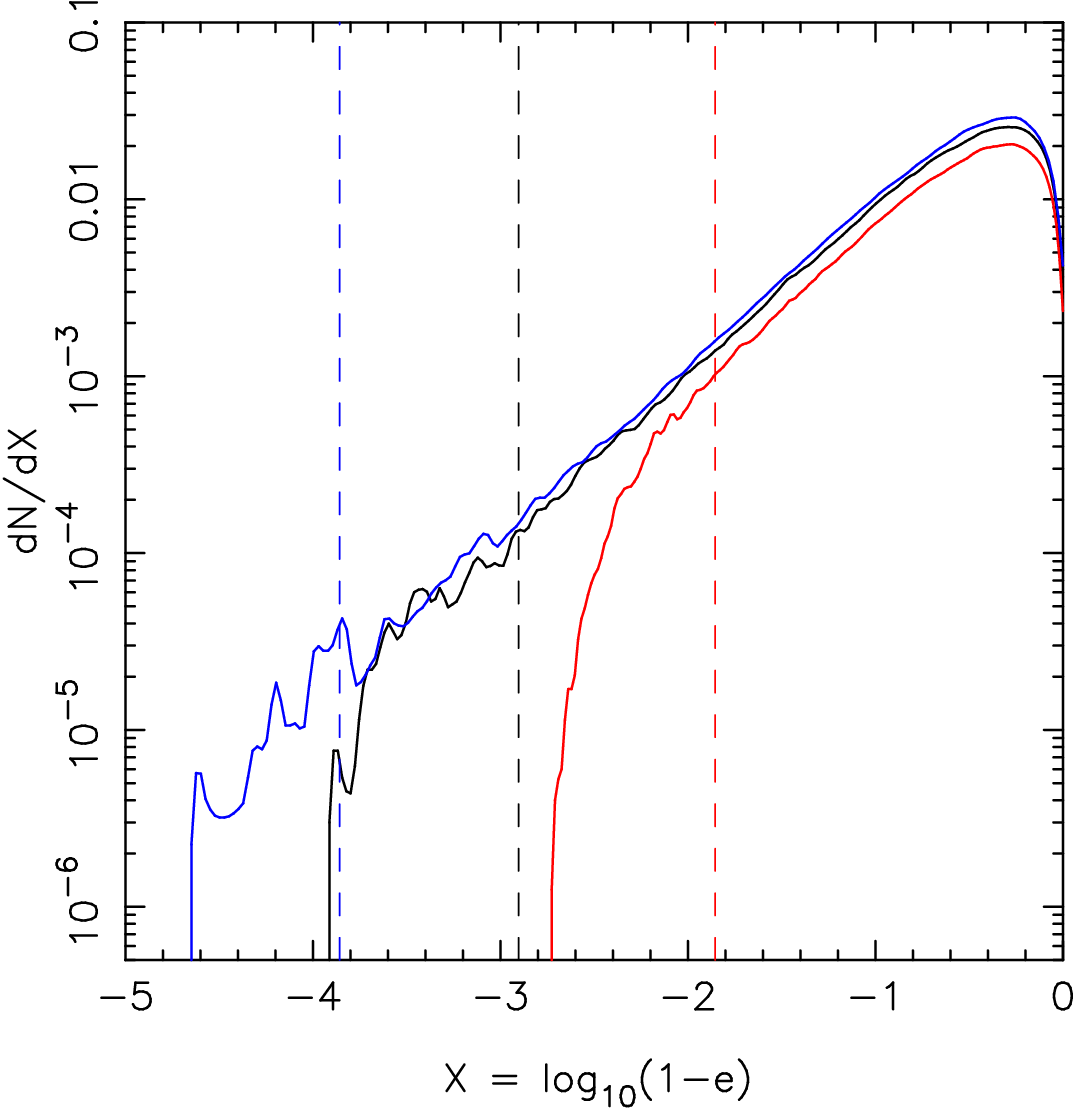}
\includegraphics[width=7.cm]{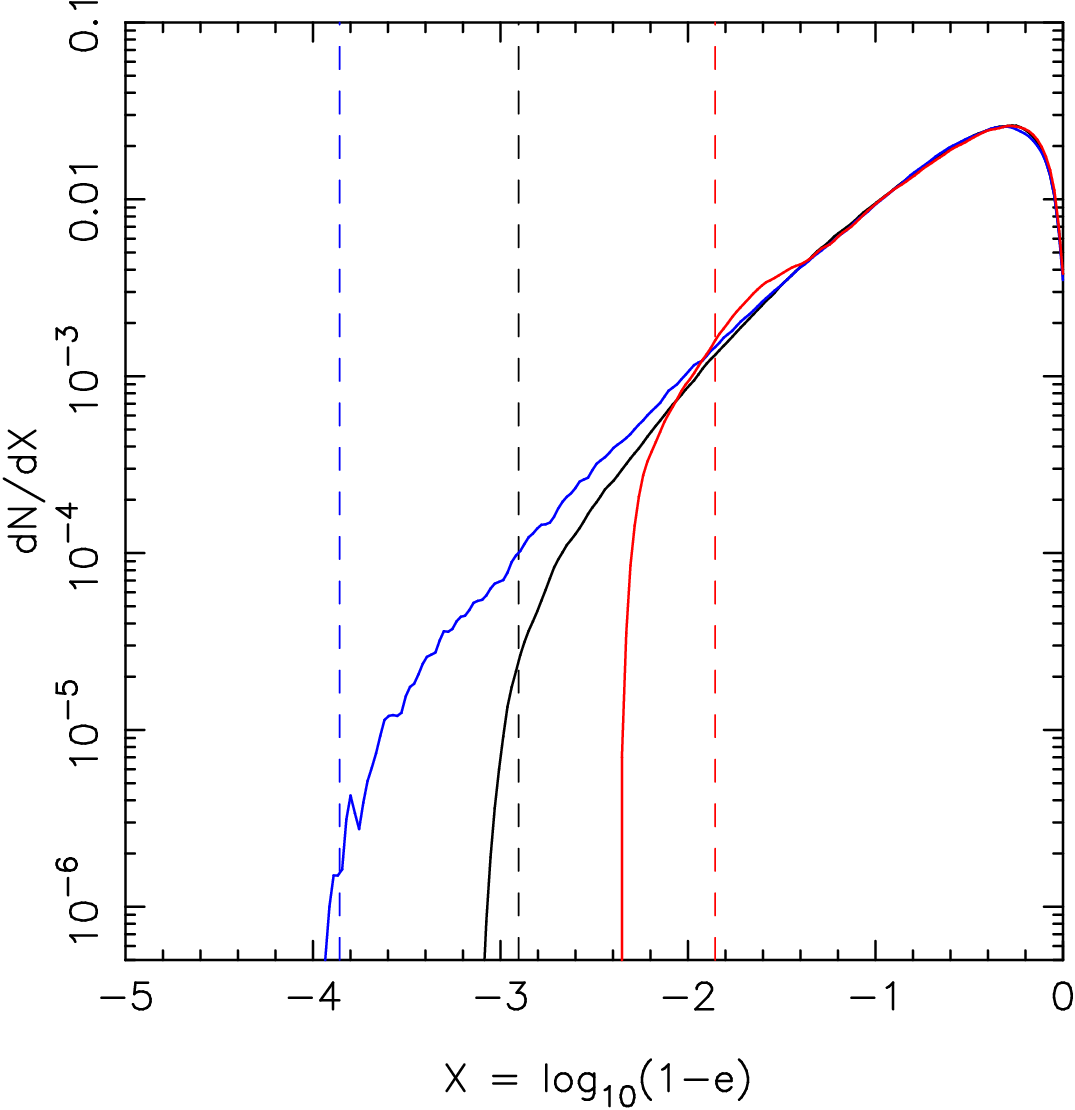}
\caption{\label{fig:mc} 
Angular momentum distributions from Monte-Carlo simulations
in which the potential was re-oriented suddenly (top)
or smoothly (bottom) each $t_\mathrm{coh}$.
{\it Red (rightmost):} $\tilde a = 2$;
{\it black:} $\tilde a = 4$;
{\it blue (leftmost):} $\tilde a = 8$.
Dashed lines show the predicted barrier location,
Eq.~(\ref{eq:lSB}).}
\end{figure}

\begin{figure}
\includegraphics[width=7.cm]{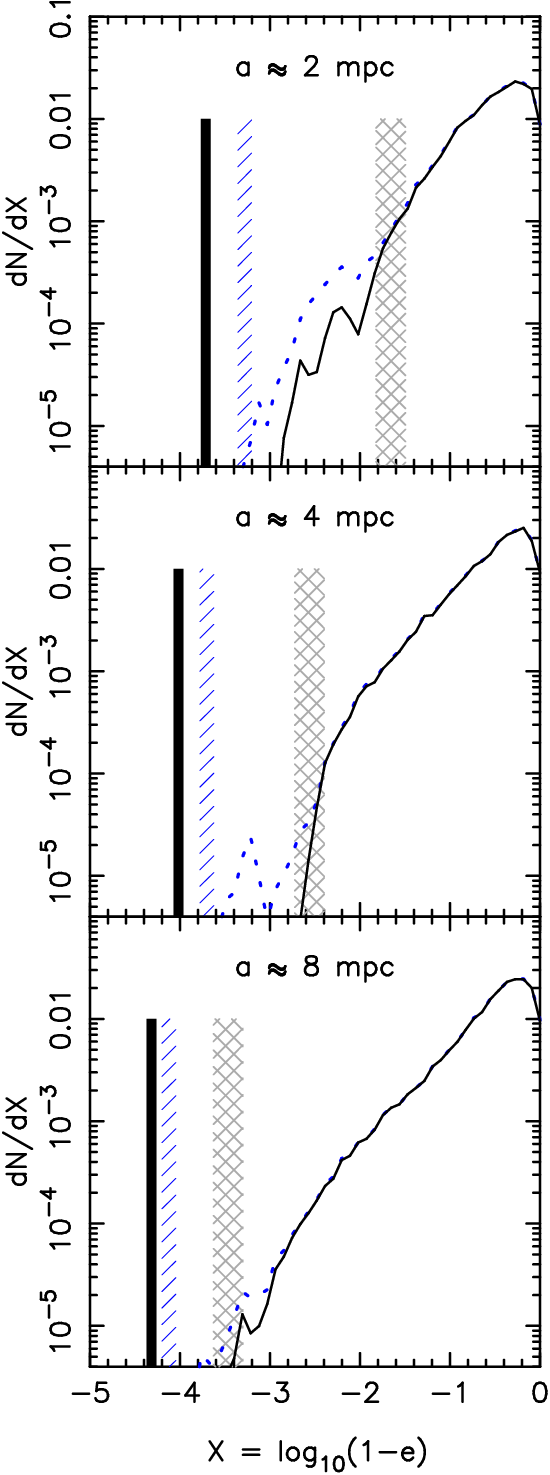}
\caption{\label{fig:dndl}
Time-averaged angular momentum distributions
of stars in the Series III $N$-body integrations, in three
intervals of semi-major axis.
The distributions were computed using all stars with
instantaneous $\tilde a$ values in a range 
$\Delta\log_{10}\tilde a=\pm 0.05$ centered on the
stated value, over the time interval $0\le t\le 2\times 10^6$ yr.
The solid (black) curves exclude stars that eventually become EMRIs;
the dotted (blue) curves include these stars.
Cross-hatched (grey) areas show the predicted location of the
Schwarzschild barrier, Eq.~(\ref{eq:SB}), given the lower
and upper limits on $\tilde a$.
Hatched (blue) areas show the capture angular momentum for EMRIs,
Eq.~(\ref{eq:cond2}).
Solid rectangles show the angular momentum at the 
assumed capture radius around the MBH.}
\end{figure}

We tested this model of barrier penetration using a Monte-Carlo code.
The 3d equations of motion~(\ref{eq:motion}) were
re-derived for an arbitary orientation of the lopsided distortion.
Starting from some randomly-chosen initial values,
an orbit was evolved in this fixed potential 
for a time $t_\mathrm{coh}$.
The orientation of the lopsided distortion was then randomized
and the integration was continued in the new potential, 
followed by another randomization of the potential etc.
In addition to the parameters 
$A_\mathrm{D}, A_\mathrm{M}$ defined in Eqs.~(\ref{eq:defAmAd})
and~(\ref{eq:params}),
this Monte-Carlo model has the additional parameter
\beq
R\equiv \nu_0 t_\mathrm{coh} = \frac{3\pi}{2A_\mathrm{D}} 
 \approx 4 {\tilde a}^{-3/2},
\eeq
the dimensionless time between potential reorientations,
where the final expression uses Eq.~(\ref{eq:paramsb}).
The number of Schwarzschild precessional periods between 
potential re-orientations is $\sim R/2\pi\langle\ell\rangle^2$.

Monte-Carlo experiments were carried out for the following 
sets of parameters:
\begin{eqnarray}
A_\mathrm{M}=7,\ \ \ A_\mathrm{D}&=&3,\ \ \ R=1.4\ \ \ (\tilde a \approx 2)\nonumber \\
A_\mathrm{M}=30,\ \ \ A_\mathrm{D}&=&10,\ \ \ R=0.5\ \ \ (\tilde a \approx 4)\nonumber \\
A_\mathrm{M}=120,\ \ \ A_\mathrm{D}&=&30,\ \ \ R=0.2\ \ \ (\tilde a \approx 8).\nonumber 
\end{eqnarray}
For each choice of parameters, 1000 Monte-Carlo experiments with
different initial seeds were carried out, and each experiment embodied
1000  re-orientations of the potential.

Fig.~\ref{fig:mc} shows the resulting, time-averaged
angular momentum distributions.
Also plotted there is the expected location of the
Schwarzschild barrier, computed using Eq.~(\ref{eq:lSB}).
While the latter is by nature approximate, Fig.~\ref{fig:mc}
reveals a tail toward low angular momenta rather than a sharp
cut-off at any value of $\ell$; orbits sometimes reach values
of $\ell$ that are $\sim$ an order of magnitude smaller than 
the predicted $\ell_\mathrm{SB}$.

The angular momentum distributions in the $N$-body 
integrations are shown in Fig.~\ref{fig:dndl}.
At all radii, there is a sharp cut-off in the distribution
at some value of $(1-e)$.
At large distances, $\tilde a=4$ and $8$, this cut-off lies
close to $\ell=\ell_\mathrm{SB}$,
while at smaller radii the distribution extends beyond the 
expected barrier location
(see also Fig.~\ref{fig:2parts}).
By comparison, while the angular momentum cut-off in the Monte-Carlo
experiments is also quite sharp, it occurs at $\ell$ values
that are somewhat lower than $\ell_\mathrm{SB}$ for all values of $a$.

The assumption that the potential changes suddenly every
$\sim t_\mathrm{coh}$ is unrealistic.
In reality, changes in the background potential are due to
the combined precession of individual orbits, which is a gradual process.
One consequence is that adiabatic invariance will be respected
for orbital actions whose conjugate angles are varying
on time scales much shorter than $t_\mathrm{coh}$.
This is not the case if the potential changes instantaneously,
as in the model just considered.
For  instance, if the period of Schwarzschild precession of an orbit
is short compared with $t_\mathrm{coh}$, its angular momentum
will be nearly conserved.
From Eq.~(\ref{eq:tratio}), this condition is satisfied for
$N$-body orbits near the barrier when $\tilde a$ is sufficiently
large; e.g. for $\tilde a=4$, $t_\mathrm{coh}/t_\mathrm{GR}\approx 60$.
For $\tilde a \le 1$, this ratio is $\lap 10$, suggesting that
adiabatic invariance will not be strictly enforced.
This is a plausible explanation for the better success of
the Monte-Carlo model at smaller radii.

To test this idea, we 
carried out a second set of Monte-Carlo experiments in which
changes in the background potential were continuous with respect to time.
The total change in the orientation of the torquing potential after each 
$t_\mathrm{coh}$ was the same as in the first set of experiments,
but now the vector describing the distortion was rotated at a fixed rate, 
along a great circle, from its initial to final orientations
during each interval.

Fig.~\ref{fig:mc} shows the results.
As expected, the angular momentum distributions are now truncated
more sharply at small values.
In the case of $\tilde a = 8$, the distribution falls 
to zero at $\ell\approx\ell_\mathrm{SB}$.
As $\tilde a$ is decreased,
the distributions extend progressively farther below the barrier,
approaching more closely to the results of the first set of 
experiments.
These distributions are quite consistent with those from the
$N$-body models, Fig.~\ref{fig:dndl}.

As noted above, the semi-empirical criterion~(\ref{eq:SB})
implies that there is a minimum value of $a$,
$\tilde a\lap C_\mathrm{SB}\approx 0.7$ 
using the adopted value of $C_\mathrm{SB}$, below which there is
no barrier.
A straightforward prediction is that stars with initial
values of the semi-major axis below $\sim 0.7$ mpc should
be able to form EMRIs, at a rate that is unaffected by the
arguments presented in this section.
We show below that this is in fact the case.

Nevertheless, a robust result of the work presented in this section
is that resonant relaxation itself is ineffective at coaxing stars
much past the Schwarzschild angular momentum barrier.
Rather, these results imply that the barrier should be
``hard,'' at least on time scales comparable with $t_\mathrm{RR}$
or $t_\mathrm{coh}$, or $10^4-10^5$ yr in these simulations.

\subsection{Barrier penetration. II.}

\begin{figure}
\includegraphics[width=8.5cm]{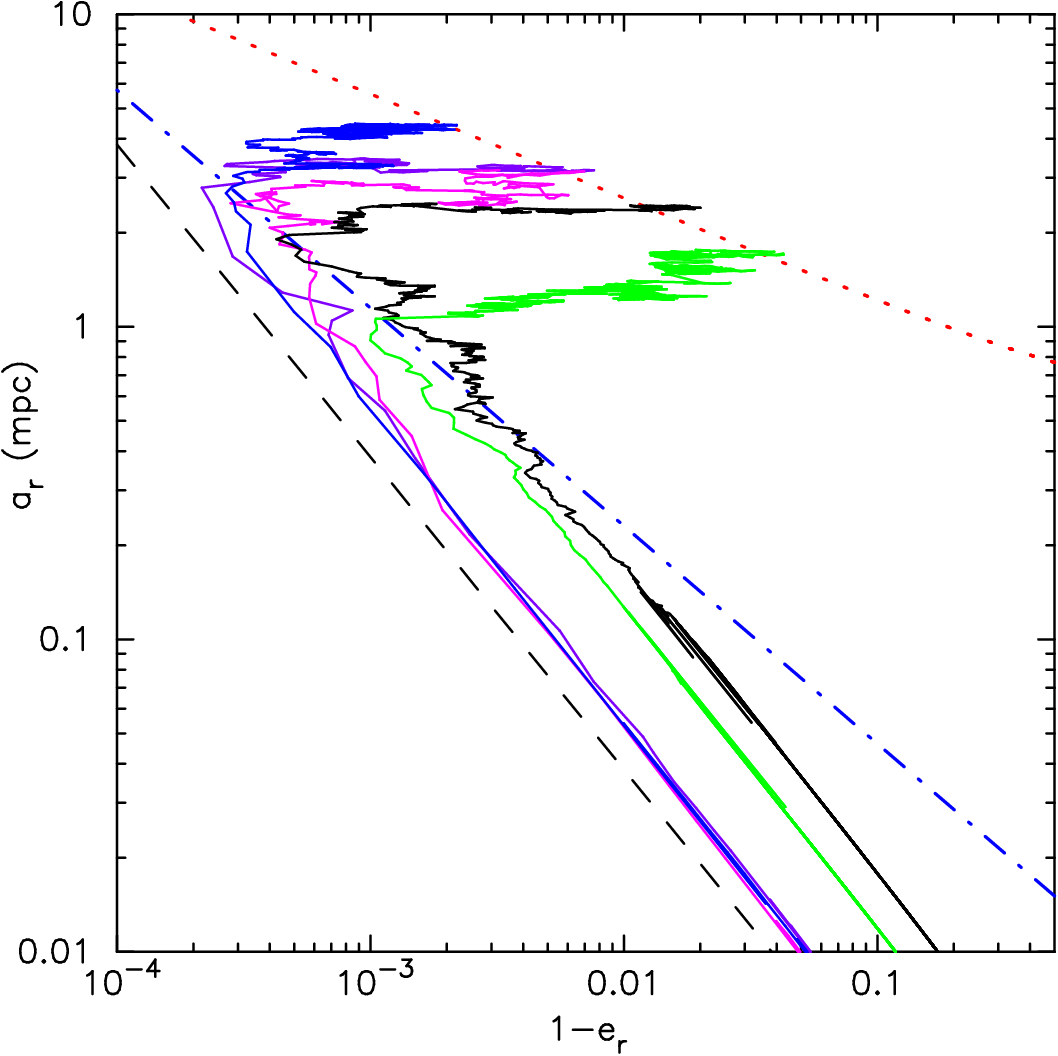}
\caption{\label{fig:gw} 
Evolutionary tracks for a subset of the stars
from Series III integrations that became EMRIs.
Dotted (red) line is the Schwarzschild barrier, Eq.~(\ref{eq:SB}).
Dash-dotted (blue) line is Eq.~(\ref{eq:cond2}), a prediction
for the critical eccentricity at which GW energy loss dominates
the evolution, assuming that the gravitational perturbations
are dominated by non-resonant relaxation.
Dashed (black) line shows the assumed capture radius.
}
\end{figure}

As noted above (cf. Fig.~\ref{fig:rates}), one or two 
stars per Myr were captured by the MBH, on average, 
in the Series III integrations, most of them as EMRIs.
Fig.~\ref{fig:gw} shows several examples.

In Newtonian systems, classical, ``two-body'' 
(non-coherent) scattering is much less effective 
than resonant relaxation at changing stars' angular momenta
when the motion is nearly Keplerian.
But this is not necessarily the case for stars on orbits
near or beyond the Schwarzschild barrier, where RR is effectively quenched
by the rapid relativistic precession.
In this section we consider the extent to which classical, 
or non-resonant (NR), relaxation can explain the EMRI events
in the $N$-body integrations.


The orbit-averaged NR relaxation time $t_\mathrm{NR}$ 
for stars of semi-major axis $a$
in our model
(a $n\propto r^{-2}$ density cusp around a MBH)
is
\begin{eqnarray}\label{eq:tNR}
&&\tnr \approx  4.6\ \mathrm{Myr}\ {\tilde a}^{1/2}
\nonumber \\
&&\left(\frac{\mh}{10^6\msun}\right)^{3/2}
\left(\frac{m}{50\msun}\right)^{-2} \left(\frac{\tilde N}{5}\right)^{-1}
\end{eqnarray}
with $\tilde N$ the number of stars within 1 mpc  (Appendix~\ref{appendC}).

Suppose that NR were the only mechanism capable of changing stars'
angular momentum leftward of the Schwarzschild barrier.
The condition for capture onto a GW-dominated orbit
would then be obtained by 
replacing $\trr$ by $\tnr$ in Eq.~(\ref{eq:tl}):
\beq\label{eq:tEMRINR}
t_\mathrm{GW} = 2(1-e)\tnr .
\eeq
We find from Eqs.~(\ref{eq:deftgw}), (\ref{eq:tNR}) 
and~(\ref{eq:tEMRINR}) the critical value of $\tilde a$
at which this condition is satisfied:
\begin{eqnarray}
\label{eq:cond2}
&&\tilde a_\mathrm{GW} = 2.0\ {\tilde r_g}^{5/7} 
\left(\frac{\mh}{m{\tilde N}\ln\Lambda}\right)^{2/7}
\left(1-e\right)^{-5/7} \\
&&\approx 9\times 10^{-3} \left(\frac{\mh}{10^6\msun}\right)
\left(\frac{m}{50\msun}\right)^{-2/7}\left(\frac{\tilde N}{5}\right)^{-2/7}
\left(1-e\right)^{-5/7}\nonumber
\end{eqnarray}
where $\tilde N$ is the number of stars with $a\le 1$ mpc.

Eq.~(\ref{eq:cond2}) is plotted as the dot-dashed (blue) line
in Fig.~\ref{fig:gw}.
After crossing this line, stars can be seen to remain near to it for
some time as their energy drops.
This diagram suggests that Eq.~(\ref{eq:cond2}) accurately specifies the region
where GW energy loss and gravitational scattering are equally
important, consistent with our assumption
that NR relaxation is the dominant mechanism for angular momentum evolution in
this part of the $(a,e)$ diagram.

Given a criterion for when a star enters the GW regime,
we can then ask how often the barrier penetration
described in the previous section would have resulted in EMRIs.

Before doing so, we note two characteristic radii associated
with $a_\mathrm{GW}$.
When
\beq
a\gap 0.8\ \mathrm{mpc} \left(\frac{\mh}{m}\right)^{1/3}
\left(\frac{\mh}{10^6\msun}\right)^{-5/8}
\tilde N^{-1/24},
\eeq
the Schwarzschild barrier lies to the left of the GW line.
In the $N$-body models considered here the critical
value is $\tilde a \approx 20$, beyond $\tilde a_\mathrm{max}=10$.
At and above such radii,
the Schwarzschild barrier would not be an impediment to EMRI
formation, and the EMRI rate (per interval of semi-major axis)
would be similar to what was found above in the Series II integrations.
This limit is probably of only academic interest however,
since in standard models of nuclei, almost all EMRIs would
originate from orbits with $a\lap 0.01$ pc.

At the other extreme in radius, the Schwarzschild barrier lies to
the right of $e=0$.
In the $N$-body models the intersection occurs at
$\tilde a = C_\mathrm{SB}\approx 0.7$; in general, 
Eqs.~(\ref{eq:barrier}) and~(\ref{eq:cond2}) give for this condition
\beq
a \lap 1.6\times 10^{-3}\mathrm{pc} \left(\frac{\mh}{m}\right)^{2/3}
\tilde N^{-1/3}.
\eeq
Since the barrier does not exist at these radii, 
the differential capture rates would also be similar
to what was observed in the Series III simulations.

For values of the semi-major axis between these two extremes
($0.7\lap \tilde a\lap 10$ in the $N$-body models),
the Schwarzschild barrier exists and lies to the right of the
critical eccentricity for GW emission.
EMRI formation at these radii requires
a substantial degree of barrier penetration.

\begin{table}
\begin{center}
\caption{Mean times to EMRI formation in the Monte-Carlo experiments}
\label{tab:tabone}
\begin{tabular}{cccccc}
\hline 
$\tilde a$  & \phantom{aaa} & $\overline{t}_1$ (yr) & 
\phantom{aaa} &$\overline{t}_2$ (yr) \\
\hline
     2 & &    $3.1\times 10^8$   & &  - \\
     4 & &   $2.8\times 10^8$   & &  - \\
     8 & &   $1.2\times 10^8$   & &  $1.5\times 10^8$ \\
\hline
\end{tabular}
\end{center}
\end{table}

We tabulated how often in the Monte-Carlo experiments from the
previous section a star passed the GW boundary.
Since not every experiment resulted in such an event,
the mean event time, in each set of experiments, 
was computed using a formula from survival analysis
\citep{Isobe-89}:
\beq
\overline{t} = \frac{1}{N_e}\sum_{i=0}^{N_e} t_i +
\frac{N_\mathrm{MC}-N_e}{N_e} T
\eeq
where $N_\mathrm{MC}$ is the total number of experiments,
$N_e$ is the number of experiments
in which the star satisfied the condition (\ref{eq:cond2})
at least once, 
$t_i$ is the time at which this first occurred,
and $T$ is the total elapsed time per experiment.

The results are presented in Table 1, for the first 
(sharp changes; $\overline t_1$)
and second (smooth changes $\overline t_2$) sets of Monte-Carlo experiments.
As expected, for large $\tilde a$, the two times are similar,
since the barrier is no impediment.
At smaller radii, the mean times are interestingly short only
in the first set of experiments; in the second set, no
events were observed.
We note a certain ``conspiracy'':
at small $a$, the degree of barrier penetration is greater, but 
the GW line lies farther from the Schwazschild barrier.

Next we consider the effectiveness of non-resonant relaxation 
at penetrating the barrier.
Several factors are relevant:

1. Because RR is so rapid to the right of the barrier,
the angular momentum distribution in this region should
remain close to that associated with an isotropic phase-space density,
i.e. $N(\ell)d\ell \approx$ constant $\times \ell d\ell$.
This contrasts with the case \cite[e.g.][]{CK-78}
where NR alone determines the 
phase space density, leading to a logarithmic decrease
in $N$ with respect to $\ell$ near the loss-cone boundary.

2. The angular momentum of a star near the barrier 
oscillates, at roughly the Schwarzschild frequency, with
amplitude $\delta\ell=\ell_+-\ell_- \approx \ell_+-\ell_\mathrm{SB}$ 
(Eq.~\ref{eq:l-l+}).
To push a star past the barrier, a NR perturbation will 
require a finite amplitude $\Delta\ell_\mathrm{NR}\gap \delta\ell$.

3. Stars remain near the barrier only for a time 
$\sim t_\mathrm{coh}$; after this, the direction of
the background torque changes, and the star random-walks
to larger angular momenta, as discussed in the previous section.

The change in $\ell$ due to NR over an interval of time
equal to $t_\mathrm{coh}$ is
\beq\label{eq:dtNR}
(\Delta\ell)_\mathrm{NR} \approx 
\left(\frac{t_\mathrm{coh}}{t_\mathrm{NR}}\right)^{1/2}.
\eeq
This change is large enough to move a star leftward
of the barrier if 
\beq\label{eq:largeenough}
(\Delta\ell)_\mathrm{NR} \gap \delta\ell = \ell_+-\ell_\mathrm{SB}
\approx \ell_+-\ell_- \approx 2\langle\ell\rangle^2A_\mathrm{D}.
\eeq
Let $\ell_\mathrm{max}(a)$ be the largest value of $\ell_+$ for
which this condition is satisfied.
Writing
\beq\label{eq:elplus}
\ell_+\approx \langle\ell\rangle + \frac{1}{2}\left(\ell_+-\ell_-\right)\approx
\langle\ell\rangle +  \langle\ell\rangle^2A_\mathrm{D}
\eeq
and eliminating $\langle\ell\rangle$ in Eqs.~(\ref{eq:largeenough})
and~(\ref{eq:elplus}) then gives
\beq\label{eq:lmax}
\ell_\mathrm{max} \approx \frac{1}{2}\left(\Delta\ell\right)_\mathrm{NR}
+ \left(2A_\mathrm{D}\right)^{-1/2}\left(\Delta\ell\right)_\mathrm{NR}^{1/2}.
\eeq

At every $a$, we expect stars with $\ell_+\lap\ell_\mathrm{max}(a)$
to be scattered leftward of the barrier in a time $t_\mathrm{coh}$.
The fraction of stars at $a$ with 
$\ell_\mathrm{SB}(a)\le\ell\le\ell_\mathrm{max}(a)$ is 
\beq
F(a) \approx \ell_\mathrm{max}^2(a) - \ell_\mathrm{SB}^2(a)
\eeq
and the time scale for stars to be lost past the barrier is
therefore
\beq
t_\mathrm{loss}(a) \equiv \left|\frac{1}{N}\frac{dN}{dt}\right|^{-1}
\approx F(a)^{-1} t_\mathrm{coh}(a).
\eeq

\begin{figure}
\includegraphics[width=8.cm]{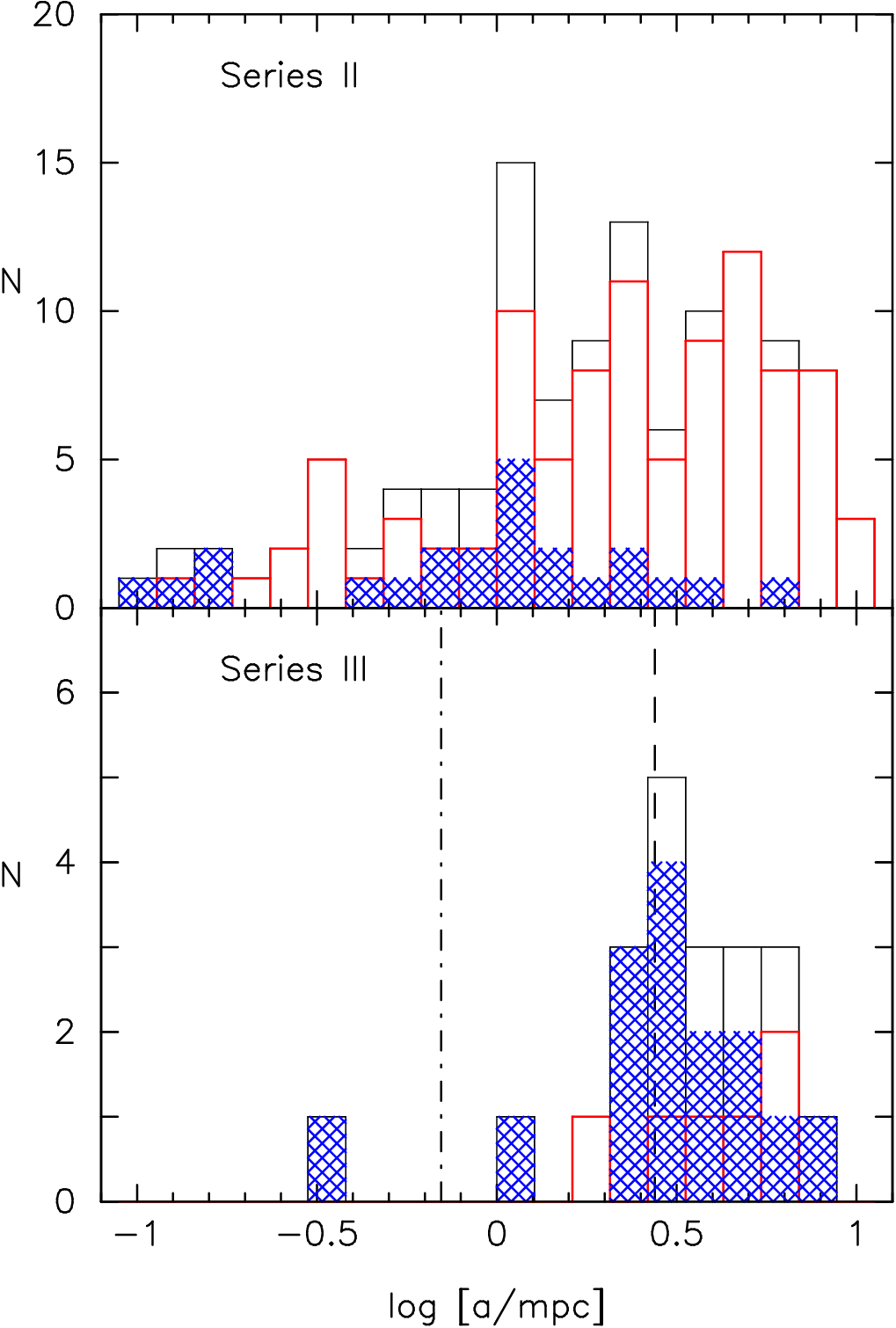}
\caption{\label{fig:twohists} 
Distribution of semi-major axes for the 
capture events from Series II (top) and Series III (bottom).
Red (unfilled) histogram shows the plunges;
blue (cross-hatched) histogram shows the EMRIs;
the total is indicated in black.
In the upper panel the initial value of $a$ is used; 
in the lower panel, the value of $a$ during the final crossing
of the Schwarzschild barrier was used.
In both panels, the elapsed time is $2\times 10^6$ yr.
To the left of the dashed vertical line in the lower panel,
non-resonant relaxation is predicted to be ineffective at pushing stars
past the Schwarzschild barrier.
To  the left of the dash-dotted vertical line,
the Schwarzschild barrier does not exist.}
\end{figure}

We evaulate $\left(\Delta\ell\right)_\mathrm{NR}$ using each of
the two choices for $t_\mathrm{coh}$ discussed above:
the mass precession time, Eq.~(\ref{eq:tM2}), which gives
\beq\label{eq:DlM}
\left(\Delta\ell\right)_\mathrm{NR,M} \approx 4.4\times 10^{-2}
\eeq
in the $N$-body models; and the vector resonant relaxation time,
Eq.~(\ref{eq:deftRRv}), for which
\beq\label{eq:Dlv}
\left(\Delta\ell\right)_\mathrm{NR,RRv} \approx 5.3\times 10^{-2}\tilde a^{1/4}.
\eeq

Tables~\ref{tab:NRtM} and~\ref{tab:NRtRRv} give the computed values of 
$F$ and $F^{-1}t_\mathrm{coh}$ for the two choices of $t_\mathrm{coh}$.
Predicted loss rates are similar for high $a$ values, and in both cases, 
NR is predicted to fail to breach the barrier when $\tilde a\lap 2-3$.

Fig.~\ref{fig:twohists} plots histograms
of the capture events in the Series III integrations.
As predicted, the number of events falls sharply for $\tilde a\lap 2-3$ mpc.
A few captures also occur from orbits with $\tilde a \lap 1$,
roughly the minimum value for which the barrier is present.

\begin{table}
\begin{center}
\caption{NR loss rates: $t_\mathrm{coh}=t_\mathrm{M}$}
\label{tab:NRtM}
\begin{tabular}{cccc}
\hline 
$\tilde a$&$t_\mathrm{coh}$ (yr)&$\ell_\mathrm{max}^2-\ell_\mathrm{SB}^2$&$t_\mathrm{loss}$ (yr) \\
\hline
     2 & $1.2\times 10^4$ & $-$ $-$ & $-$ $-$  \\
     3 & $1.6\times 10^4$ & $4.9\times 10^{-5}$ & 
         $3.1\times 10^8$ \\
     4 & $1.8\times 10^4$ & $2.1\times 10^{-3}$ & 
         $8.5\times 10^6$ \\
     5 & $2.0\times 10^4$ & $2.5\times 10^{-3}$ & 
         $8.1\times 10^6$ \\
     6 & $2.2\times 10^4$ & $2.5\times 10^{-3}$ & 
         $8.9\times 10^6$ \\
     8 & $2.5\times 10^4$ & $2.2\times 10^{-3}$ & 
         $1.2\times 10^7$ \\
     10 &$2.8\times 10^4$ & $1.9\times 10^{-3}$ & 
         $1.5\times 10^7$ \\
\hline
\end{tabular}
\end{center}
\end{table}

\begin{table}
\begin{center}
\caption{NR loss rates: $t_\mathrm{coh}=t_\mathrm{RR,v}$}
\label{tab:NRtRRv}
\begin{tabular}{cccc}
\hline 
$\tilde a$&$t_\mathrm{coh}$ (yr)&$\ell_\mathrm{max}^2-\ell_\mathrm{SB}^2$&$t_\mathrm{loss}$ (yr) \\
\hline
     2 & $2.6\times 10^4$ & $-$ $-$ & $-$ $-$  \\
     3 & $3.9\times 10^4$ & $5.6\times 10^{-3}$ & 
         $7.2\times 10^6$ \\
     4 & $5.2\times 10^4$ & $7.2\times 10^{-3}$ & 
         $7.4\times 10^6$ \\
     5 & $6.5\times 10^4$ & $7.5\times 10^{-3}$ & 
         $9.0\times 10^6$ \\
     6 & $7.8\times 10^4$ & $7.3\times 10^{-3}$ & 
         $1.1\times 10^7$ \\
     8 & $1.4\times 10^5$ & $6.9\times 10^{-3}$ & 
         $1.5\times 10^7$ \\
     10 & $1.3\times 10^5$ & $6.6\times 10^{-3}$ & 
         $2.0\times 10^7$ \\
\hline
\end{tabular}
\end{center}
\end{table}

We can also compare predicted and measured event rates.
From Fig.~\ref{fig:rates}, the mean capture rate at early times
in the Series III integrations is $\sim 1-2\times 10^{-6}$ yr$^{-1}$.
(Four out of 19 of the events were associated with orbits
below the Schwarzschild barrier, reducing the mean rate of
barrier-crossing events slightly.)
The number of stars initially with $\tilde a\gap 2-3$ is
$\sim 35-40$, and the loss times in Table~2 are roughly
$1\times 10^7$ yr at these radii.
The predicted event rate is therefore $3-4\times 10^{-6}$ yr$^{-1}$
-- in reasonable agreement with the measured values given the
crudeness of the model. 
We note that our model can be expected to overestimate the 
capture rate since it ignores the possibility of a star returning
to the right of the barrier after crossing it.

We will present a more detailed calculation of the barrier
penetration rate due to NR in a later paper
\cite{AM-11}.

\section{\label{sec:discuss}Discussion}

Here we discuss briefly how the key results from the $N$-body 
experiments can be extended to nuclear star clusters with more
general properties.
We treat this topic in more detail in 
Papers II and III \cite{AM-11,ABM-11}; in particular,
we do not attempt here to derive absolute EMRI rate estimates
for general clusters.

We begin by collecting some of the important relations derived
above and expressing them in more general form.

Combining the parameter dependence of Eq.~(\ref{eq:barrier}) 
with the empirical normalization of Eq.~(\ref{eq:SB}), we find
for the angular momentum that defines the Schwarzschild barrier:
\begin{widetext}
\begin{eqnarray}\label{eq:SBd}
\left(1-e^2\right)_\mathrm{SB} &\approx&
1.9 \left(\frac{C_\mathrm{SB}}{0.7}\right)^2
\left(\frac{r_g}{a}\right)^2 
\left(\frac{\mh}{m}\right)^2 
\frac{1}{N}\nonumber\\
&\approx& 0.23 \left(\frac{C_\mathrm{SB}}{0.7}\right)^2
\left(\frac{a}{\mathrm{mpc}}\right)^{-2}
\left(\frac{\mh}{10^6\msun}\right)^{4}
\left(\frac{m}{10\msun}\right)^{-2}
\left(\frac{N}{10^2}\right)^{-1};
\end{eqnarray}
here and below, $N$ is the number of stars within radius $a$.
$N(a)\propto a$ was not assumed in deriving this expression.
However, that assumption {\it was} made in deriving 
Eq.~(\ref{eq:cond2}), the condition that GW emission dominate
stellar encounters.
We can generalize that relation to a cluster with
arbitrary density profile using the approximate scaling of the
NR relaxation time:
\beq
t_\mathrm{NR} \propto \frac{\mh^2}{m^2} \frac{P_r}{N}
\eeq
\cite[e.g.][]{HA-06b}, together with the exact, orbit-averaged
expression for $t_\mathrm{NR}$ in the case $n(r)\propto r^{-2}$,
Eq.~(\ref{eq:tNR}), to write
\beq
t_\mathrm{NR}\approx 6\ \mathrm{Myr} 
\left(\frac{a}{\mathrm{mpc}}\right)^{3/2}
\left(\frac{\mh}{10^6\msun}\right)^{3/2}
\left(\frac{m}{10\msun}\right)^{-2}
\left(\frac{N}{10^2}\right)^{-1}.
\eeq
The condition for GW emission to dominate relaxation then becomes
\beq \label{eq:GWcond}
\left(1-e^2\right)_\mathrm{GW}
\approx 2\times 10^{-3} 
\left(\frac{a}{\mathrm{mpc}}\right)^{-1}
\left(\frac{\mh}{10^6\msun}\right)^{7/5}
\left(\frac{m}{10\msun}\right)^{-2/5}
\left(\frac{N}{10^2}\right)^{-2/5}.
\eeq
If physical capture by the MBH is assumed to occur when $r\le\rm$,
$\rm=\Theta r_g$, then the critical eccentricity for capture is
\beq
\left(1-e^2\right)_\mathrm{capt} \approx 8\times 10^{-4}
\left(\frac{\Theta}{8}\right)
\left(\frac{a}{\mathrm{mpc}}\right)^{-1}
\left(\frac{\mh}{10^6\msun}\right).
\eeq

We define $a_\mathrm{plunge}$ to be the value of $a$ such
that $\left(1-e^2\right)_\mathrm{capt} = 
\left(1-e^2\right)_\mathrm{GW}$;
at larger $a$, all stars plunge.
We find that $a_\mathrm{plunge}$ is defined implicitly by
\beq
N(<a_\mathrm{plunge}) \approx 1.1\times 10^3\left(\frac{\Theta}{8}\right)^{-5/2}
\left(\frac{M_\bullet}{10^6\msun}\right)
\left(\frac{m}{10\msun}\right)^{-1}.
\eeq

\begin{figure*}
\includegraphics[width=10.cm,angle=-90.]{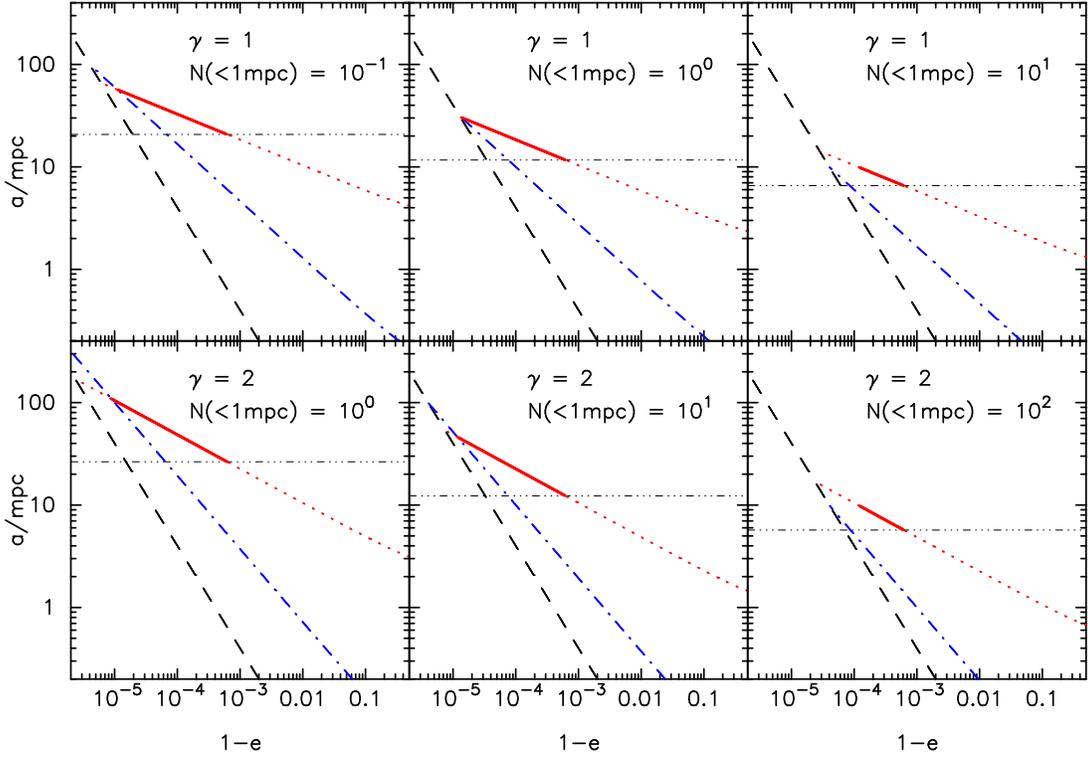}
\caption{\label{fig:cases} 
Illustrating the critical curves defined in the text
for nuclear star clusters obeying density laws
$n(r)\propto r^{-\gamma}$, with various slopes and normalizations.
{\it Dashed (black) line:} capture radius  ($r_p=8r_g$);
{\it dash-dotted (blue) line:} radius at which GW emission
dominates stellar perturbations, eq.~(\ref{eq:GWcond}).
The Schwarzschild barrier, eq.~(\ref{eq:SBd}), is shown as the red line;
it is solid where conditions allow EMRI formation.
Below the horizontal line, non-resonant relaxation is
expected to be inefficient at pushing stars past the barrier
(eqs.~~\ref{eq:ula}-\ref{eq:ulb},~\ref{eq:impl}.)
$\mh=10^6\msun$ and $m=10\msun$ were assumed.
}
\end{figure*}

The Schwarzschild barrier intersects the GW line when
\beq\label{eq:ula}
\left(\frac{a}{\mathrm{mpc}}\right)
\left(\frac{N}{10^2}\right)^{3/5} \approx 
120 \left(\frac{C_\mathrm{SB}}{0.7}\right)^2
\left(\frac{\mh}{10^6\msun}\right)^{13/5}
\left(\frac{m}{10\msun}\right)^{-8/5}
\eeq
and it intersects the capture line when
\beq\label{eq:ulb}
\left(\frac{a}{\mathrm{mpc}}\right)
\left(\frac{N}{10^2}\right)\approx 
300 \left(\frac{C_\mathrm{SB}}{0.7}\right)^2
\left(\frac{\Theta}{8}\right)^{-1}
\left(\frac{\mh}{10^6\msun}\right)^{3}
\left(\frac{m}{10\msun}\right)^{-2}.
\eeq
One of these two relations defines the effective upper
limit to the radial extent of the Schwarzschild barrier.
Setting $e=0$ in Eq.~(\ref{eq:SBd}) gives the lower radial limit:
\beq
\left(\frac{a}{\mathrm{mpc}}\right)^2
\left(\frac{N}{10^2}\right)\approx 
0.2 \left(\frac{C_\mathrm{SB}}{0.7}\right)^2
\left(\frac{\mh}{10^6\msun}\right)^{4}
\left(\frac{m}{10\msun}\right)^{-2}.
\eeq

Another key parameter is the minimum value of $a$ for
which non-resonant relaxation is able to penetrate the
Schwarzschild barrier (Sect. Vc).
Combining Eqs.~(\ref{eq:tM2}), (\ref{eq:dtNR}), (\ref{eq:lmax}), 
(\ref{eq:lSB}), and~(\ref{eq:SBd}),
we find that the critical value of $a$ satisfies the 
implicit relation
\beq\label{eq:impl}
\left(\frac{a}{\mathrm{mpc}}\right)_\mathrm{penetrate} 
\approx
15 \left(\frac{C_\mathrm{SB}}{0.7}\right)
\left(\frac{\mh}{10^6\msun}\right)^{5/2}
\left(\frac{m}{10\msun}\right)^{-3/2}
\left(\frac{N}{10^2}\right)^{-1/2}.
\eeq
In deriving this expression, we have equated $t_\mathrm{coh}$ 
with $t_\mathrm{M}$.

We apply these expressions to nuclear star clusters obeying
\beq
n(r) = n_0 r^{-\gamma}
\eeq
i.e.
\beq\label{eq:Nlta}
N\equiv N(<a) = N_{<1} \tilde a^{3-\gamma}
\eeq
where $\tilde a \equiv a/\mathrm{mpc}$ and $N_{<1}$
is the number of stars with $a\le 1$ mpc.
Combining Eqs.~(\ref{eq:Nlta}) and (\ref{eq:impl}), we find
\beq
\tilde a_\mathrm{penetrate} \approx
\left[140\left(\frac{C_\mathrm{SB}}{0.7}\right)
\left(\frac{\mh}{10^6\msun}\right)^{5/2}
\left(\frac{m}{10\msun}\right)^{-3/2}
N_{<1}^{-1/2}\right]^{2/(5-\gamma)}.
\eeq
\end{widetext}

Fig.~\ref{fig:cases} plots the relations defined above
for clusters of $m=10\msun$ BHs around a $\mh=10^6\msun$ MBH.
We chose likely values of $\gamma$ and $N_{<1}$ 
following the discussion in Ref.~\cite{MAMW-10}.
As in the $N$-body models, there is generally a rather small
range of $a$ values from which EMRIs can form: small enough
that GW emission can overcome stellar perturbations, but
large enough that non-resonant relaxation can push stars past
the Schwarzschild barrier.
Interestingly, for sufficiently dense clusters, this range can
go to zero, implying essentially no EMRIs; however it appears
that the required densities are one or two orders of magnitude 
larger than expected for real galactic nuclei \cite{MAMW-10}.

\section{\label{sec:conclude}Conclusions}

1. $N$-body integrations have been used, for the
first time, to directly simulate the long-term evolution of
relativistic clusters of compact stars around 
massive black holes (MBHs), both Schwarzschild and Kerr, 
and to compute the rate of extreme-mass-ratio inspirals
(EMRIs).

2. When relativistic terms are omitted from the equations
of motion, stars are scattered into the MBH
at rates that are in good agreement with those expected
from the theory of resonant relaxation (RR).

3. Relativistic precession suppresses RR,
leading to an effectively maximum value of the eccentricity 
at each value of the semi-major axis.
This ``Schwarzschild barrier'' strongly inhibits EMRI
formation, leading to capture rates that are factors
$\sim 10-100$ lower than in the non-relativistic case.

4. We use an approximate Hamiltonian formulation of the perturbed
equations of motion to explore two possible mechanisms for barrier 
penetration: one related to resonant relaxation 
and the other to non-resonant relaxation (NR).
We show that NR is effective at penetrating the Schwarzschild
barrier only for orbits with semi-major axes above a certain value,
and this prediction is verified in the $N$-body integrations.
Approximate expressions for the capture rate are derived
and shown to be consistent with the rates observed in the simulations.

\bigskip
\begin{acknowledgments}
DM was supported in part by the National Science Foundation under  
grants no. AST 08-07910, 08-21141 and by the National Aeronautics and 
Space Administration under grant no. NNX-07AH15G.
TA was supported by ERC Starting Grant 202996, and by
DIP-BMBF grant 71-0460-0101.
CMW was supported in part by the National Science Foundation under  
grant no. PHY 06-52448 and 09–65133, 
the National Aeronautics and Space Administration, grant no. NNG-
06GI60G, and the Centre National de la Recherche Scientifique, Programme Internationale de
la Coop\'eration Scientifique (CNRS-PICS), grant no. 4396. 
Parts of this research were carried out while CMW was a visitor to 
the Institut d’Astrophysique de Paris, and while DM, SM and CMW were visitors to
the Benoziyo Center for Astrophysics, Weizmann Institute of Science; 
the hospitality of these two institutes is
gratefully acknowledged.
We thank P. Amaro-Seoane, B. Baror, and P. Saha for useful discussions.
\end{acknowledgments}

\appendix
\begin{widetext}
\section{Hamiltonian model}
\label{appendB}

Here we use standard techniques \cite[e.g.][]{MV-10}
to derive the equations describing the rates of change of the 
Keplerian (osculating) elements of a star
moving in the potential~(\ref{eq:Potential}):
\begin{equation}
\Phi({\bf r}) = -\frac{G\mh}{r} + \Phi_p, \ \ 
\Phi_p=\Phi_s \ln\left(\frac{r}{r_o}\right) - Sa\cos\theta
\label{eq:PotentialA}
\end{equation}
and including the time-averaged 
effects of Schwarzschild precession.

We begin by transforming from Cartesian coordinates to
Delaunay variables \citep[e.g.][]{Goldstein-02}
which are action-angle variables  in the Kepler problem.
The Delaunay action variables are the radial action $I=(G\mh a)^{1/2}$, 
the angular momentum $L$, and the projection of $\mathbf{L}$ onto
the $z$ axis $L_z$.
The conjugate angle variables are the mean anomaly $w$,
the argument of the periapse $\omega$, 
and the longitude of the ascending node $\Omega$.
In the Keplerian case, five of these are constants;
the exception is $w$ which increases linearly with time at a rate
\begin{equation}  
\nu_r=(GM_\bullet)^2/I^3 .
\label{Equation:nu_rA}
\end{equation}

The Hamiltonian, averaged over $w$, is
\begin{subequations}
\begin{eqnarray}
\overline{\cal H} &=& -\frac{1}{2}\left(\frac{GM_\bullet}{I}\right)^2 + 
\overline\Phi_p, \\\label{eq:defPhiAvA}
\overline\Phi_p &\equiv& \oint\frac{dw}{2\pi} \Phi_p 
= \frac{1}{2\pi}\int_0^{2\pi} dE\,(1-e\cos E)\,\Phi_p({\bf r}).
\end{eqnarray}
\end{subequations}
In the final term, $E$ is the eccentric anomaly,
where $r=a(1-e\cos E)$ and the eccentricity is  $e=\sqrt{1-L^2/I^2}$.
After the averaging, $\overline{\cal H}$ is independent of $w$,
and $I$ is conserved, as is the semi-major axis $a$.
We are left with four variables and with $\overline\Phi_p$ 
as  the effective Hamiltonian of the system. 

The orbit-averaged Hamiltonian describes slow, precessional dynamics.
Superposed on the slow variations described by the averaged
dynamics are fast oscillations, with frequencies 
$\sim (G\mh/a^3)^{1/2}$  and with fractional amplitudes
$\delta\approx a\left(\Phi-\overline{\Phi}\right)/G\mh$.
If $\delta\ll 1$, i.e. if $\mh\gg M_\star$,
we can ignore these fast oscillations.

After expressing the Cartesian coordinates in terms of the
Delaunay variables,
the results of the averaging are
\begin{subequations}
\label{eq:PhiAvA}
\begin{eqnarray}
\label{eq:PhiAvaA}
\overline{\Phi}_p &=& \overline\Phi_M + \overline\Phi_D + 
\overline\Phi_\mathrm{GR}, \\ 
\overline\Phi_M &=& \frac{GM(a)}{a}\left[C(a) + F(e)\right],\\
C(a) &=& \ln\left(\frac{a}{r_0}\right)+1-\ln 2 , \ \ \ \ 
F(e) = \ln\left(1+\sqrt{1-e^2}\right) - \sqrt{1-e^2},\\
\overline\Phi_D &=&Sae\sin i \sin\omega,\\
\overline\Phi_\mathrm{GR} &=& -\frac{3G^2\mh^2}{c^2a^2}\left(1-e^2\right)^{-1/2}.
\end{eqnarray}
\end{subequations}
The averaged dipole potential, $\overline\Phi_D$, is expressed in terms of 
the orbital inclination $i$ where
$\cos i = L_z/L$; $i=0$ for an orbit that is 
perpendicular to  the major axis of the dipole.
The last term, $\overline\Phi_\mathrm{GR}$, reproduces
the orbit-averaged rate of Schwarzschild periapse advance,
Eq.~(\ref{eq:nuGR}).
The longitude of the ascending node, $\Omega$, does not appear
due to symmetry of the potential about the $z$-axis.

In the limit $e\rightarrow 1$, $F(e)\rightarrow -\ell^2/2 $.

We define a dimensionless time $\tau=\nu_0t$ where
\beq\label{eq:defnu0A}
\nu_0 = \nu_r \frac{3G\mh}{c^2a},
\eeq
the Schwarzschild precession frequency in the limit $e\rightarrow 0$.
Dropping constant terms 
(including terms that depend only on semi-major axis $a$),
the dimensionless Hamiltonian describing the perturbed motion becomes
\beq
\label{eq:hamil}
H \equiv \frac{\overline\Phi_p}{\nu_0I} =
- \left(1-e^2\right)^{-1/2} 
+ A_M F(e) +  A_D e \sin i \sin \omega, 
\eeq
with $A_\mathrm{M}, A_\mathrm{D}$ defined in Eq.~(\ref{eq:defAmAd}).
The equations of motion,
\beq
\frac{\partial\omega}{\partial\tau}=\frac{\partial H}{\partial\ell},\ \ 
\frac{\partial\ell}{\partial\tau}=-\frac{\partial H}{\partial\omega},\ \ 
\frac{\partial\Omega}{\partial\tau}=\frac{\partial H}{\partial\ell_z},\ \ 
\frac{\partial\ell_z}{\partial\tau}=-\frac{\partial H}{\partial\Omega}=0
\eeq
are given explicitly in Eqs.~(\ref{eq:motion}).

\section{Non-resonant relaxation}
\label{appendC}

Here we summarize the orbit-averaged equations describing
changes in angular momentum due to non-resonant relaxation (NR) 
and derive the angular-momentum diffusion coefficient for the
$N$-body models \citep[e.g.][]{LS-77,MT-1999,MM-2003}.

In terms of the binding energy per unit mass $E=-v^2/2+\psi(r)=G\mh/2a$, 
where $\psi(r)=G\mh/r$, and the normalized angular momentum
$R\equiv L^2/L_c^2=\ell^2=1-e^2$,
the Fokker-Planck equation describing diffusion in
angular momentum due to NR is 
\beq
\frac{\partial N}{\partial t}=
\frac{1}{2}\frac{\partial}{\partial R}
\left[\langle\left(\Delta R\right)^2
\rangle\frac{\partial N}{\partial R}\right]
\eeq
where
\beq
\label{eq:NER}
N(E,R)dEdR=N(a,e)dade
\eeq
is the number density of stars in (energy, angular momentum) space,
and $\langle\left(\Delta R\right)^2\rangle$ 
is the diffusion coefficient
in $R$, i.e. the sum, over a unit
interval of time, of $(\Delta R)^2$ due to encounters.

Taking the limit $R\rightarrow 0$
and averaging over one orbital period, this becomes
\beq
\label{eq:diffusion}
\frac{\partial N}{\partial t} = {\bar\mu} \frac{\partial }{\partial R}
\left(R \frac{\partial N}{\partial R}\right)
\eeq
where ${\bar\mu}(E)$ is the orbit-averaged diffusion coefficient:
\beq
{\overline\mu}(E)\equiv 
P_r(E)^{-1}\oint \frac{dr}{v_r}
\lim_{R\rightarrow 0}
\frac{\langle(\Delta R)^2\rangle}{2R}
\label{eq:defq}
\eeq
and the integral is over one full radial period.
$\overline{\mu}(E)$ 
is precisely the orbit average of the
inverse angular momentum relaxation time defined by
Hopman \& Alexander \cite{HA-05} and henceforth we write
$\overline\mu^{-1}\equiv \tnr$.

Let $f(E)$ be the phase-space number density of stars;
it is related to $N(E)$ by 
\begin{eqnarray}
\label{eq:fofN}
f(E) &=& \frac{1}{\sqrt{2}\pi^3}\left(G\mh\right)^{-3}
E^{5/2} N(E) \\
&=& f_0E^{\gamma-3/2}.
\end{eqnarray}
The latter expression assumes $n(r) \propto r^{-\gamma}$;
the $N$-body models have $\gamma=2$.
The local diffusion cofficient is expressible in terms of $f(E)$ via
\begin{subequations}
\begin{eqnarray}
\lim_{R\rightarrow 0} \frac{\langle(\Delta R)^2\rangle}{2R} &=&
\frac{32\pi^2r^2G^2m^2\ln\Lambda}{3L_c^2}\left(3I_{1/2} - I_{3/2} + 2I_0\right),\\
I_0(E) &=& \int_0^E f(E') dE',\\
I_{n/2}(E,r) &=& \left\{2\left[\psi(r)-E\right]\right\}^{-n/2}
\int_E^{\psi}\left\{2\left[\psi(r)-E'\right]\right\}^{n/2} f(E') dE'
\end{eqnarray}
\end{subequations}
where $\ln\Lambda\approx\ln[\mh/(2m)]$ is the Coulomb logarithm;
in the $N$-body models $\ln\Lambda\approx 9$.

The orbit averages are
\beq
\overline{I}(E) = \frac{1}{\sqrt{2}}\int_0^{G\mh/E}
\frac{r^2 dr}{\sqrt{\Psi-E}} I(E,r);
\eeq
setting $\gamma=2$, the value in the $N$-body models, we find
\begin{subequations}
\begin{eqnarray}
\overline{I}_0(E) &=& \frac{5\sqrt{2}\pi}{48}f_0
E^{-2} \left(G\mh\right)^3,\\
\overline{I}_{1/2}(E) &=& \frac{\pi}{16\sqrt{2}} \left(\ln 16 - 2\right)f_0
E^{-2} \left(G\mh\right)^3,\\
\overline{I}_{3/2}(E) &=& \frac{\pi}{96\sqrt{2}} \left(11-12\ln 2\right)
f_0E^{-2} \left(G\mh\right)^3
\end{eqnarray}
\end{subequations}
and
\begin{subequations}
\begin{eqnarray}
3\overline{I}_{1/2}-\overline{I}_{3/2}+2\overline{I}_0 &=&
C_\gamma f_0E^{-2} \left(G\mh\right)^3, \\
C_2 &=& \frac{7\pi\sqrt{2}}{192}\left(12\ln 2 - 1\right)\\
&\approx& 1.18533
\end{eqnarray}
\end{subequations}
so that
\beq
\overline{\mu}(E) = C_2\frac{64\pi\sqrt{2}}{3} G^2m^2\ln\Lambda f(E).
\eeq
We note that $\overline{\mu}(E)\propto f(E)$, a result that
holds for arbitrary $\gamma$.

Eqs.~(\ref{eq:Nofae}, \ref{eq:NER}, \ref{eq:fofN}) combine to give
$f$ in terms of $a$:
\beq
f(a) = \frac{1}{4\pi^3} \left(G\mh\right)^{-3/2} N_0 a^{-1/2}
\eeq 
where $N_0=r^{-1}N(<r)=a^{-1}N(<a)$.
Then
\beq
\overline{\mu}^{-1}(a) \equiv \tnr(a) = 
C_2^{-1}\frac{3\sqrt{2}\pi^2}{32}
\frac{\left(G\mh\right)^{3/2}}
{G^2m^2 N_0\ln\Lambda}a^{1/2}.
\eeq

\end{widetext}

\bibliography{biblio}

\end{document}